\documentclass[showpacs,preprintnumbers,amsmath,amssymb]{revtex4}

\usepackage{epsfig}% Include figure files
\usepackage{graphicx}% Include figure files
\usepackage{dcolumn}% Align table columns on decimal point
\usepackage{bm}% bold math

\begin{document}

%%%%%%%%%%%%%%%%%%%%%%%%%%%%%%%%%%%%%%%%%%%%%%%%%%%%%%%%%%%%%%%%%%%%%%%%%%
%                               Title                                    %
%%%%%%%%%%%%%%%%%%%%%%%%%%%%%%%%%%%%%%%%%%%%%%%%%%%%%%%%%%%%%%%%%%%%%%%%%%

\title{Finite-Energy Landau Liquid Theory for the 1D Hubbard Model:\\
Pseudoparticle Energy Bands and Degree of
Localization/Delocalization}
\author{J. M. P. Carmelo}
\affiliation{GCEP-Center of Physics, University of Minho, Campus
Gualtar, P-4710-057 Braga, Portugal}
\author{P. D. Sacramento}
\affiliation{Departamento de F\'{\i}sica and CFIF, Instituto
Superior T\'ecnico, P-1049-001 Lisboa, Portugal}
\date{22 July 2002}
%\date{\today}

%%%%%%%%%%%%%%%%%%%%%%%%%%%%%%%%%%%%%%%%%%%%%%%%%%%%%%%%%%%%%%%%%%%%%%%%%%
%                              abstract                                  %
%%%%%%%%%%%%%%%%%%%%%%%%%%%%%%%%%%%%%%%%%%%%%%%%%%%%%%%%%%%%%%%%%%%%%%%%%%

\begin{abstract}
In this paper we consider the one-dimensional Hubbard model and
study the deviations from the ground state values of double
occupation which result from creation or annihilation of holons,
spinons, and pseudoparticles. These quantum objects are such that
all energy eigenstates are described by their occupancy
configurations. The band-momentum dependence of the obtained
double-occupation spectra provides important information on the
degree of localization/delocalization of the real-space lattice
electron site distribution configurations associated with the
pseudoparticles. We also study the band-momentum, on-site
electronic repulsion, and electronic density dependence of the
pseudoparticle energy bands. The shape of these bands plays an
important role in the finite-energy spectral properties of the
model. Such a shape defines the form of the lines in the
momentum-energy/frequency plane where the peaks and edges of the
one-electron and two-electron spectral weight of physical
operators are located. Our findings are useful for the study of
the one-electron and two-electron spectral weight distribution of
physical operators.
\end{abstract}


\maketitle
%%%%%%%%%%%%%%%%%%%%%%%%%%%%%%%%%%%%%%%%%%%%%%%%%%%%%%%%%%%%%%%%%%%%%%%%%%
%                              body of paper                             %
%%%%%%%%%%%%%%%%%%%%%%%%%%%%%%%%%%%%%%%%%%%%%%%%%%%%%%%%%%%%%%%%%%%%%%%%%%
%%%%%%%%%%%%%%%%%%%%%%%%%%%%%%%%%%%%%%%%%%%%%%%%%%%%%%%%%%%%%%%%
\section{INTRODUCTION}

Recently, there has been a renewed experimental interest in the
exotic one-electron and two-electron spectral properties of
quasi-one-dimensional materials
\cite{Hussey,Menzel,Fuji,Hasan,Ralph,Gweon,Monceau,Takenaka,Mizokawa,Moser,Mihaly,Vescoli00,Denlinger,Fujisawa,Kobayashi,Bourbonnais,Vescoli,Zwick,Mori,Kim}.
Some of these experimental studies observed unusual
finite-energy/frequency spectral properties, which are far from
being well understood. For values of the excitation energy larger
than the transfer integrals associated with electron hopping
between the chains, the one-dimensional (1D) Hubbard model
\cite{Lieb,Takahashi,Martins98,Rasetti,Hubbard} is expected to
provide a good description of the physics of these materials
\cite{Hasan,Ralph,Vescoli,Mori}. Moreover, recent angle-resolved
ultraviolet photoemission spectroscopy revealed very similar
spectral fingerprints from both high-$T_c$ superconductors and
quasi-one-dimensional compounds \cite{Menzel}. The similarity of
the ultraviolet data for these two different systems could be
evidence of the occurrence of a charge-spin separation associated
with holons and spinons. The anomalous temperature dependence of
the spectral function could also indicate a dimensional crossover
\cite{Menzel,Granath,Orgard,Carlson}. The results of Refs.
\cite{Zaanen,Antonio} also suggest that the unconventional
spectral properties observed in two-dimensional (2D) materials
could have a 1D origin. Thus the holons and spinons could play an
important role in spectral properties of both 1D and 2D
low-dimensional materials.

The present study is related the to both the investigations of
Refs. \cite{I} and \cite{III}. In reference \cite{I} the
non-perturbative organization of the $N$ electrons and
$N^h=[2N_a-N]$ electronic holes, which results from the
many-electron interactions, was studied. In that paper it was
found that all energy eigenstates of the model can be described in
terms of occupancy configurations of $\sigma_c=\pm 1/2$ holons,
$\sigma_s=\pm 1/2$ spinons, and $c$ pseudoparticles where
$\sigma_c$ and $\sigma_s$ are $\eta$-spin and spin projections
respectively. Moreover, the $c,\nu$ pseudoparticles (and $s,\nu$
pseudoparticles) associated with Takahasi's charge (and spin)
ideal string excitations of length $\nu$ \cite{Takahashi,I} were
found to be $\eta$-spin singlet $2\nu$-holon (and spin singlet
$2\nu$-spinon) composite quantum objects. (The $c$ and $\alpha
,\nu$ pseudoparticles are obtained naturally and directly from
analysis of the Bethe-ansatz solution \cite{I}.) In that
reference, the $\pm 1/2$ holons (and $\pm 1/2$ spinons) which are
not part of such $2\nu$-holon composite $c,\nu$ pseudoparticles
(and $2\nu$-spinon composite $s,\nu$ pseudoparticles) were called
$\pm 1/2$ Yang holons (and $\pm 1/2$ HL spinons). In the
designation {\it Yang holon} (and {\it HL spinon}) Yang refers to
C. N. Yang (HL stands for Heilmann and Lieb) \cite{I}. Following
the notation of Ref. \cite{I}, here we call $N_c$ and
$N_{\alpha,\,\nu}$ the numbers of $c$ pseudoparticles and
$\alpha,\nu$ pseudoparticles belonging to branches $\alpha =c,\,s$
and $\nu=1,2,3,...$ respectively. On the other hand, it was found
in Ref. \cite{I} that the charge excitations associated with the
$c$ pseudoparticles are independent of the $\eta$-spin-holon and
spin-spinon excitations. Moreover, the results of Ref. \cite{III}
reveal that the above quantum objects can be expressed in terms of
{\it rotated electrons}. According to these results the first step
of the non-perturbative diagonalization of the model corresponds
to a mere unitary rotation in Hilbert space which maps the
electrons onto rotated electrons. For rotated electrons double
occupation is a good quantum number for all values of the on-site
repulsion. Interestingly, such a rotation corresponds to a unitary
transformation previously introduced by Harris and Lange
\cite{Harris}. Such a transformation is shown in Ref. \cite{III}
to play a key role in the expression of the holon and spinon
number operators in terms of electronic operators for the whole
parameter space of the model. The $N_c$ $c$ pseudoparticles (and
$N_c$ spinons) describe the charge part (and spin part) of the
$N_c$ rotated electrons which singly occupy lattice sites. Thus
the charge and spin degrees of freedom of such a
$N_c$-rotated-electron fluid separate. For each $c$ pseudoparticle
there is a chargeon \cite{I} and a rotated-electronic hole. Such a
chargeon corresponds to the charge part of a rotated electron
which singly occupies a site. We note that the numbers $N_c$,
$[N-N_c]/2$, and $[N^h-N_c]/2$ equal the number of
rotated-electron singly occupied sites, doubly occupied sites, and
empty sites respectively. The non-perturbative organization of the
electronic degrees of freedom also includes a
$[N-N_c]$-rotated-electron fluid associated with the
rotated-electron pairs occupying $[N-N_c]/2$ sites. Each such a
doubly occupied site corresponds to a spin singlet rotated
electron pair which is nothing but a $-1/2$ holon. Such a
two-fluid picture shows some formal analogies to those which
characterize a superconductor or a superfluid. Finally, the
rotated-electron $[N^h-N_c]/2$ empty sites are nothing but the
$+1/2$ holons.

The main goal of this paper is the study of the $c$
pseudoparticle, $\alpha ,\nu$ pseudoparticle, $-1/2$ Yang holon,
and $-1/2$ HL spinon double-occupation spectra. These functions
are evaluated by use of the Hellmann-Feynman theorem
\cite{Carmelo86,Carmelo88}. In the case of the pseudoparticles we
find band-momentum dependent double-occupation spectra. Analysis
of these spectra provides important information about the
localization/deslocalization degree of the real-space lattice
electronic occupancy configurations which describe the
pseudoparticles. We find that for some of the pseudoparticle
branches such a degree of localization/delocalization is strongly
dependent on the value of the pseudoparticle band momentum. Our
analysis leads to a double-occupation functional whose
coefficients are the double-occupation spectra. This study
confirms that the complex electron site distribution
configurations of the real-space lattice which describe the energy
eigenstates are dependent on the value of $U/t$. We also consider
$U/t\rightarrow\infty$ electron double-occupation selection rules.
These rules are used in Ref. \cite{III} in finding exact
rotated-electron selection rules which provide for all values of
the on-site electronic repulsion the number of holons and spinons
of the final states which contribute most significantly to
few-electron correlation functions. These selection rules imply
that a large number of matrix elements between energy eigenstates
are vanishing. This simplifies the derivation of general
expressions for correlation functions at finite excitation energy.
This problem is studied in Refs. \cite{IV,V}.

In addition, we study the band-momentum, on-site electronic
repulsion $U$, and electronic density $n$ dependence of the
pseudoparticle energy bands. Such pseudoparticle energy bands are
the coefficients of the first-order term of the finite-energy
Landau-liquid functional. That functional was obtained in Ref.
\cite{Carmelo97} by expanding the $c$ pseudoparticle, $c,\nu$
pseudoparticle, and $s,\nu$ pseudoparticle band-momentum
distribution functions around their ground-state values. This is
similar to the case of the quasi-particle energy functional of
Fermi liquid theory \cite{Pines,Baym}. The pseudoparticle bands
are the basic blocks of the energy spectra of the elementary
excitations of the 1D Hubbard model, as for example the ones
studied in Refs.
\cite{Ovchinnikov,Coll,Woynarovich,Choy,Klumper,Deguchi}.

The finite-energy theory constructed here and in Refs.
\cite{I,III,IV,V} is applied elsewhere to the study of the
finite-energy/frequency one-electron and two-electron spectral
properties of quasi-one dimensional materials. Reference
\cite{optical} is an application of a preliminary version of our
finite-energy theory. The evaluation of the spectral functions of
the model reveals that the one-electron and two-electron spectral
weight of physical operators contains peaks and/or edges which are
located on well-defined lines in the momentum $k$ and
energy/frequency $\omega$ plane. Importantly, the shape of these
peak and edge lines is fully defined by the band-momentum
dependence of the pseudoparticle energy bands studied in this
paper. Moreover, the expressions of the finite-energy correlation
functions involve these pseudoparticle bands \cite{V}. Therefore,
our investigation is important for the study of the one-electron
and two-electron spectral-weight distribution of physical
operators. The study of the localization-delocalization degree of
the real-space lattice electronic occupancy configurations which
describe the pseudoparticles also contributes to important
information about the physics behind these elementary quantum
objects. Such a study also provides information about the electron
site distribution configurations of the real-space lattice which
describe the energy eigenstates.

The paper is organized as follows: In Sec. II we introduce the 1D
Hubbard model. The concepts of CPHS ensemble space and subspace
are introduced in Sec. III. Here CPHS stands for {\it $c$
pseudoparticle, holon, spinon}. In Sec. IV the double occupation
functional and its pseudoparticle, $-1/2$ Yang holon, and $-1/2$
HL spinon spectra are obtained and discussed. In this section we
also consider $U/t\rightarrow\infty$ selection rules in the values
of double occupation deviations which result from
$\cal{N}$-electron excitations. In Sec. V we study the
band-momentum dependence of the energy pseudoparticle bands.
Finally, the concluding remarks are presented in Sec. VI.

%%%%%%%%%%%%%%%%%%%%%%%%%%%%%%%%%%%%%%%%%%%%%%%%%%%%%%%%%%%%%%%%%%%%%%%%%%
\section{THE 1D HUBBARD MODEL}

In a chemical potential $\mu $ and magnetic field $H$ the 1D
Hubbard Hamiltonian can be written as,

\begin{equation}
\hat{H}={\hat{H}}_{SO(4)} + \sum_{\alpha =c,\,s}\mu_{\alpha
}\,{\hat{S}}^{\alpha}_z
\label{H}
\end{equation}
where the Hamiltonian,

\begin{equation}
{\hat{H}}_{SO(4)} = {\hat{H}}_{H} - (U/2)\,\hat{N} + (U/4)\,N_a \,
,
\label{HSO4}
\end{equation}
has $SO(4)$ symmetry \cite{HL,Yang89} and

\begin{equation}
{\hat{H}}_{H} = \hat{T}+U\,\hat{D} \, ,
\label{HH}
\end{equation}
is the basic 1D Hubbard model. On the right-hand side of Eq.
(\ref{HH}),

\begin{equation}
\hat{T}=-t\sum_{j,\,\sigma}[c_{j,\,\sigma}^{\dag}\,c_{j+1,\,\sigma}
+ h. c.] \, ,
\label{Top}
\end{equation}
is the kinetic-energy operator, $t$ is the first-neighbor transfer
integral, and $U$ is the on-site repulsion associated with
electronic double occupation. The operator

\begin{equation}
\hat{D} = \sum_{j}
\hat{n}_{j,\,\uparrow}\,\hat{n}_{j,\,\downarrow} \, ,
\label{Dop}
\end{equation}
counts the number of electron doubly-occupied sites and the
operator

\begin{equation}
\hat{n}_{j,\,\sigma}= c_{j,\,\sigma }^{\dagger }\,c_{j,\,\sigma }
\, ,
\label{njsig}
\end{equation}
counts the number of spin $\sigma$ electrons at site $j$. The
operators $c_{j,\,\sigma }^{\dagger }$ and $c_{j,\,\sigma}$ which
appear in the above equations are the spin $\sigma $ electron
creation and annihilation operators at site $j$ respectively.
Moreover, on the right-hand side of Eq. (\ref{H}), $\mu_c=2\mu$,
$\mu_s=2\mu_0 H$, $\mu_0$ is the Bohr magneton, and

\begin{equation}
{\hat{S }}^c_z=
-{1\over 2}[N_a-\hat{N}] \, ; \hspace{1cm}
{\hat{S }}^s_z=
-{1\over 2}[{\hat{N}}_{\uparrow}-{\hat{N}}_{\downarrow}]
\, ,
\label{Sz}
\end{equation}
are the diagonal generators of the $SU(2)$ $\eta$-spin $S^c$ and
spin $S^s$ algebras \cite{HL,Yang89} respectively. On the
right-hand side of Eqs. (\ref{HSO4}) and (\ref{Sz}) the number of
lattice sites $N_a$ is even and large, $N_a/2$ is odd,
${\hat{N}}=\sum_{\sigma} \hat{N}_{\sigma}$, and
${\hat{N}}_{\sigma}=\sum_{j} \hat{n}_{j,\sigma}$.

There are $N_{\uparrow}$ spin-up electrons and $N_{\downarrow}$
spin-down electrons in the chain of $N_a$ sites and with lattice
constant $a$ associated with the model (\ref{H}). We assume
periodic boundary conditions and employ units such that $a=\hbar
=1$. When $N_{\sigma }$ is odd the Fermi momenta are given by
$k_{F\sigma }^{\pm }=\pm \left[k_{F\sigma } -{\pi\over
{N_a}}\right]$ where

\begin{equation}
k_{F\sigma }={\pi N_{\sigma }\over {N_a}} \, .
\label{kFs}
\end{equation}
When $N_{\sigma }$ is even the Fermi momenta are given by
$k_{F\sigma}^{+} = k_{F\sigma}$ and $k_{F\sigma}^{-} = -
\left[k_{F\sigma }-{2\pi\over {N_a}}\right]$ or by
$k_{F\sigma}^{+} = \left[k_{F\sigma } -{2\pi\over {N_a}}\right]$
and $k_{F\sigma}^{-} = - k_{F\sigma }$. Often we can ignore the
$1/N_a$ corrections of these expressions and consider
$k_{F\sigma}^{\pm}\simeq\pm k_{F\sigma}=\pm \pi n_{\sigma }$ and
$k_F=[k_{F\uparrow}+ k_{F\downarrow}]/2=\pi n/2$, where
$n_{\sigma}=N_{\sigma}/N_a$ and $n=N/N_a$. The electronic density
reads $n=n_{\uparrow }+n_{\downarrow}$, and the spin density is
given by $m=n_{\uparrow}-n_{\downarrow}$. This paper uses the same
notations as Ref. \cite{I} and often uses and refers to the
equations introduced in that paper. For instance $M_{c,\,\pm 1/2}$
(and $M_{s,\,\pm 1/2}$) denotes the number of $\pm 1/2$ holons
(and $\pm 1/2$ spinons), whereas $L_{c,\,\pm 1/2}$ (and
$L_{s,\,\pm 1/2}$) denotes the number of $\pm 1/2$ Yang holons
(and $\pm 1/2$ HL spinons).

The Hamiltonian $\hat{H}_{SO(4)}$ defined in Eq. (\ref{HSO4})
commutes with the six generators of the $\eta$-spin $S_c$ and spin
$S_s$ algebras \cite{HL,Yang89}, their off-diagonal generators
reading,

\begin{equation}
{\hat{S}}^c_{+}=\sum_{j}(-1)^j c_{j,\,\downarrow}^{\dag}\,
c_{j,\,\uparrow}^{\dag} \, ; \hspace{1cm}
{\hat{S}}^c_{-}=\sum_{j}(-1)^j
c_{j,\,\uparrow}\,c_{j,\,\downarrow} \, ,
\label{Sc}
\end{equation}
and

\begin{equation}
{\hat{S}}^s_{+}= \sum_{j}
c_{j,\,\downarrow}^{\dag}\,c_{j,\,\uparrow} \, ; \hspace{1cm}
{\hat{S}}^s_{-}=\sum_{j}
c_{j,\,\uparrow}^{\dag}\,c_{j,\,\downarrow} \, .
\label{Ss}
\end{equation}

The Bethe-ansatz solvability of the 1D Hubbard model is restricted
to the Hilbert subspace spanned by the lowest-weight states (LWSs)
of the $\eta$ spin and spin algebras, {\it i.e.} such that
$S^{\alpha}= -S^{\alpha}_z$ \cite{Essler}.

%%%%%%%%%%%%%%%%%%%%%%%%%%%%%%%%%%%%%%%%%%%%%%%%%%%%%%%%%%%%%%%%
\section{CPHS-ENSEMBLE SPACES AND GROUND-STATE DISTRIBUTIONS AND RAPIDITIES}

The studies of Ref. \cite{I} considered Hilbert subspaces
associated with fixed values of $\eta$-spin $S_c$, spin $S_s$,
holon number $M_c$, and spinon number $M_s$. That was the choice
suitable for the investigations on $\eta$-spin-holon, spin-spinon,
and charge $c$ pseudoparticle separation of the excitations of the
1D Hubbard model at all energy scales. However, fixing $S_c$,
$S_s$, $M_c$, $M_s$ does not fix the $N_{\uparrow}$ and
$N_{\downarrow}$ electron numbers.

In later sections of this paper and in Refs. \cite{III,IV,V}
transitions from a ground state with given spin $\sigma$ electron
numbers to excited states are studied. Thus it is convenient for
these studies to consider subspaces with fixed number of spin
$\sigma$ electrons. From the use of Eq. (\ref{Sz}) and Eq. (35) of
Ref. \cite{I}, the $N_{\uparrow}$ and $N_{\downarrow}$ electron
numbers can be expressed in terms of the $M_{c,\,+1/2}$ and
$M_{c,\,-1/2}$ holon numbers and $M_{s,\,+1/2}$ and $
M_{s,\,-1/2}$ spinon numbers as follows,

\begin{equation}
N_{\uparrow} = {1\over 2}\Bigl\{N_a - (M_{c,\,+1/2} -
M_{c,\,-1/2}) + (M_{s,\,+1/2} - M_{s,\,-1/2})\Bigr\} \, ,
\label{NupM}
\end{equation}
and

\begin{equation}
N_{\downarrow} = {1\over 2}\Bigl\{N_a - (M_{c,\,+1/2} -
M_{c,\,-1/2}) - (M_{s,\,+1/2} - M_{s,\,-1/2})\Bigr\} \, ,
\label{NdoM}
\end{equation}
respectively. We find from the use of Eq. (32) of Ref. \cite{I}
that the $M_{c,\,+1/2}$ holon and $M_{s,\,+1/2}$ spinon numbers
are exclusive functions of the $N_c$ $c$-pseudoparticle number and
$M_{c,\,-1/2}$ holon and $M_{s,\,-1/2}$ spinon numbers and can be
written as,

\begin{equation}
M_{c,\,+1/2} = N_a - N_c - M_{c,\,-1/2} \, ; \hspace{1cm}
M_{s,\,+1/2} = N_c - M_{s,\,-1/2}
\label{M+NcM-} \, .
\end{equation}

It is convenient for our studies to express the $N_{\uparrow}$ and
$N_{\downarrow}$ electron numbers in terms of the $N_c$ $c$
pseudoparticle numbers, $M_{c,\,-1/2}$ holon numbers, and $
M_{s,\,-1/2}$ spinon numbers, with the result,

\begin{equation}
N_{\uparrow} = N_c + M_{c,\,-1/2} - M_{s,\,-1/2} \, ,
\label{NupNM}
\end{equation}
and

\begin{equation}
N_{\downarrow} = M_{c,\,-1/2} + M_{s,\,-1/2} \, ,
\label{NdoNM}
\end{equation}
respectively.

We call an {\it electron ensemble space} a Hilbert subspace
spanned by all states with fixed values for the $N_{\uparrow}$ and
$N_{\downarrow}$ electron numbers. Furthermore, we call a {\it
CPHS ensemble space}, where CPHS stands for $c$ pseudoparticle,
$-1/2$ holon, and $-1/2$ spinon, a Hilbert subspace spanned by all
states with fixed values for the numbers $N_c$, $M_{c,\,-1/2}$,
and $M_{s,\,-1/2}$ of $c$ pseudoparticles, $-1/2$ holons, and
$-1/2$ spinons respectively. It follows from Eqs. (\ref{NupNM})
and (\ref{NdoNM}) that in general, an electron ensemble space
contains several CPHS ensemble spaces. This means that different
choices of the numbers $N_c$, $M_{c,\,-1/2}$, and $M_{s,\,-1/2}$
can lead to the same electron numbers $N_{\uparrow}$ and
$N_{\downarrow}$. In contrast, according to Eqs. (\ref{NupNM}) and
(\ref{NdoNM}), a given choice of $N_c$, $M_{c,\,-1/2}$, and
$M_{s,\,-1/2}$ always corresponds to an unique choice for the
$N_{\uparrow}$ and $N_{\downarrow}$ values of the electron
numbers.

Each CPHS ensemble space contains several subspaces with different
numbers $L_{c,\,-1/2}$ of $-1/2$ Yang holons, $L_{s,\,-1/2}$ of
$-1/2$ HL spinons, and sets of numbers $\{N_{c,\,\nu}\}$ and
$\{N_{s,\,\nu}\}$ of $c,\,\nu$ and $s,\,\nu$ pseudoparticles
respectively, corresponding to different $\nu=1,2,3,...$ branches.
According to the results of Ref. \cite{I}, the fixed values of
each subspace obey the constraints imposed by the following
equations,

\begin{equation}
M_{c,\,-1/2} = L_{c,\,-1/2}+ \sum_{\nu =1}^{\infty} \nu \,
N_{c,\,\nu}  \, ; \hspace{1cm} M_{s,\,-1/2} = L_{s,\,-1/2}+
\sum_{\nu =1}^{\infty} \nu \, N_{s,\,\nu}  \, . \label{MLNcs}
\end{equation}

We call {\it CPHS ensemble subspace} a Hilbert subspace spanned by
all states with fixed values for the numbers $N_c$,
$L_{c,\,-1/2}$, and $L_{s,\,-1/2}$ and for the sets of numbers
$\{N_{c,\,\nu}\}$ and $\{N_{s,\,\nu}\}$ corresponding to
$\nu=1,2,3,...$ branches. It follows from Eqs. (\ref{MLNcs}) that
in general, a CPHS ensemble space contains several CPHS ensemble
subspaces. This means that different choices of the values for
$L_{c,\,-1/2}$ and $L_{s,\,-1/2}$ and sets of numbers
$\{N_{c,\,\nu}\}$ and $\{N_{s,\,\nu}\}$ corresponding to
$\nu=1,2,3,...$ branches can lead to the same values for
$M_{c,\,-1/2}$ and $M_{s,\,-1/2}$. In contrast, a given choice of
the values of $N_c$, $L_{c,\,-1/2}$, and $L_{s,\,-1/2}$ and sets
$\{N_{c,\,\nu}\}$ and $\{N_{s,\,\nu}\}$ corresponding to
$\nu=1,2,3,...$ branches always corresponds to an unique choice of
the values of $N_c$, $M_{c,\,-1/2}$, and $M_{s,\,-1/2}$.

Finally, we note that one does not need to provide the values of
$M_{c,\,+1/2}$ and $M_{s,\,+1/2}$ in order to specify a CPHS
ensemble space. These numbers are given by Eq. (\ref{M+NcM-}) and
are not independent. Furthermore, from combination of Eqs. (30),
(32), and (34) of Ref. \cite{I} one finds that

\begin{eqnarray}
L_{c,\,+1/2} & = & N_a - N_c - 2\sum_{\nu =1}^{\infty}
\nu\,N_{c,\,\nu} - L_{c,\,-1/2} \, ; \nonumber \\
L_{s,\,+1/2} & = & N_c - 2\sum_{\nu =1}^{\infty} \nu\,N_{s,\,\nu}
- L_{s,\,-1/2} \, .
\label{ScsNN}
\end{eqnarray}
Thus $L_{c,\,+1/2}$ and $L_{s,\,+1/2}$ are not independent and one
does not need to provide these values in order to specify a CPHS
ensemble subspace. Therefore, often we do not consider the values
of the holon numbers $M_{c,\,+1/2}$ and $L_{c,\,+1/2}$ and of the
spinon numbers $M_{s,\,+1/2}$ and $L_{s,\,+1/2}$ in the
expressions considered below.

According to the studies of Ref. \cite{I}, our choice of using the
numbers $N_c$ and $M_{c,\,-1/2}$ to label the charge sector of a
CPHS ensemble space corresponds to a description of charge
transport in terms of electrons. In this case the elementary
charge carriers are the $N_c$ chargeons and $M_{c,\,-1/2}$ $-1/2$
holons, which carry charge $-e$ and $-2e$ respectively. Here $-e$
denotes the charge of the electron. The alternative use of the
numbers $N_c$ and $M_{c,\,+1/2}$ to define the charge sector of a
CPHS ensemble space would correspond to a description of charge
transport in terms of electronic holes. In this case the
elementary charge carriers would be the $N_c$ rotated electronic
holes and $M_{c,\,+1/2}$ $+1/2$ holons, which carry charge $+e$
and $+2e$ respectively \cite{I}. Also the transport of spin has
two alternative descriptions which correspond to the $-1/2$
spinons and $+1/2$ spinons respectively. Here we have chosen to
label the spin sector of the CPHS ensemble spaces by the number
$M_{s,\,-1/2}$ of $-1/2$ spinons.

The finite-energy Landau liquid functional discussed in Ref.
\cite{I} and used in the ensuing sections corresponds to the CPHS
ensemble subspace description in terms of the values of the
numbers of $-1/2$ Yang holons and $-1/2$ HL spinons and of the
values of the numbers of $c$ pseudoparticles at band-momentum $q$
and of $\alpha ,\nu$ pseudoparticles belonging to $\alpha =c,s$
and $\nu =1,2,3,...$ branches at band-momentum $q$. According to
the studies of Ref. \cite{I}, such a choice corresponds again to
the description of charge transport in terms of electrons. In this
case the $c,\nu$ pseudoparticle carries charge $-2\nu\,e$.

%%%%%%%%%%%%%%%%%%%%%%%%%%%%%%%%%%%%%%%%%%%%%%%%%%%%%%%%%%%%%%%%
\section{ELECTRON DOUBLE OCCUPATION OF THE PSEUDOPARTICLES, YANG HOLONS, AND HL SPINONS}

Let us denote the number of doubly occupied sites by $D$. By
number of doubly occupied sites we mean here the expectation value
of the operator (\ref{Dop}),

\begin{equation}
D\equiv \langle\hat{D}\rangle = \sum_{j}\langle
c_{j,\,\uparrow}^{\dag}\,c_{j,\,\uparrow}\,
c_{j,\,\downarrow}^{\dag}\,c_{j,\,\downarrow}\rangle \, .
\label{D}
\end{equation}

Another useful quantity is the kinetic energy $T$, which we define
as the expectation value of the operator (\ref{Top}),

\begin{equation}
T\equiv -t\sum_{j,\,\sigma}[\langle
c_{j,\,\sigma}^{\dag}\,c_{j+1,\,\sigma}\rangle + \langle
c_{j+1,\,\sigma}^{\dag}\,c_{j,\,\sigma}\rangle ] \, .
\label{T}
\end{equation}

Let $\vert\psi\rangle$ be an energy eigenstate of the Hamiltonian
(\ref{HH}) of energy
$E_H=\langle\psi\vert{\hat{H}}_{H}\vert\psi\rangle$ given by the
general expression,

\begin{eqnarray}
E_H & = & -2t {N_a\over 2\pi} \int_{q_c^{-}}^{q_c^{+}} dq\, N_c
(q) \cos k(q)
\nonumber \\
& + & 4t {N_a\over 2\pi}\sum_{\nu=1}^{\infty} \int_{q^{-}_{c,\,\nu
}}^{q^{+}_{c,\,\nu}} dq\, N_{c,\,\nu} (q)\,{\rm Re}\,\Bigl\{
\sqrt{1 - (\Lambda_{c,\,\nu} (q) + i \nu U/4t)^2}\Bigr\}
+U\,L_{c,\,-1/2} \, ,
\label{EH}
\end{eqnarray}
where $k(q)$ and $\Lambda_{c,\,\nu} (q)$ are the rapidity
functionals defined by the integral Eqs. (13)-(16) of Ref.
\cite{I} and $N_c (q)$ and $N_{c,\,\nu} (q)$ are pseudoparticle
band-momentum distribution functions also defined therein. The
expressions of the band-momentum limiting values of the integrals
on the right-hand side of Eq. (\ref{EH}) are given in Eqs.
(B14)-(B17) of Ref. \cite{I}. This energy expression is obtained
by combining the energy expression provided by the Bethe-ansatz
solution \cite{Takahashi,Carmelo97,Deguchi} with the $SO(4)$
symmetry of the model and the holon and spinon description
introduced in Ref. \cite{I}. Note that such a general energy
expression refers to the whole Hilbert space of the 1D Hubbard
model. The term involving the number of $-1/2$ Yang holons
vanishes in the case of the LWSs of the $SU(2)$ $\eta$-spin
algebra associated with the Bethe-ansatz solution and is not
provided by such a solution.

According to the Hellmann-Feynman theorem
\cite{Carmelo86,Carmelo88}, the electron double-occupation $D$,
and kinetic energy $T$, associated with the above energy
eigenstate are given by

\begin{equation}
D \equiv \langle\psi\vert{\partial{\hat{H}}_{H}\over\partial U}
\vert\psi\rangle ={\partial
\langle\psi\vert{\hat{H}}_{H}\vert\psi\rangle \over\partial U} =
{\partial E_{H}\over\partial U} \, ,
\label{Dee}
\end{equation}
and

\begin{equation}
T \equiv \langle\psi\vert t{\partial{\hat{H}}_{H}\over\partial t}
\vert\psi\rangle =t {\partial
\langle\psi\vert{\hat{H}}_{H}\vert\psi\rangle \over\partial t} = t
{\partial E_{H}\over\partial t} \, ,
\label{Tee}
\end{equation}
respectively, where ${\hat{H}}_{H}$ is the Hamiltonian defined in
Eq. (\ref{HH}) and the energy $E_H$ is given by Eq. (\ref{EH}).

The use of expression (\ref{Dee}) in the evaluation of the
electron double-occupation spectrum associated with creation of a
$c$ pseudoparticle, $\alpha,\nu$ pseudoparticle, $-1/2$ Yang
holon, or $-1/2$ HL spinon  involves the pseudoparticle Landau
liquid energy functional. This functional was introduced in Ref.
\cite{Carmelo97} and expressed in terms of holon and spinon
numbers in Ref. \cite{I}.

%%%%%%%%%%%%%%%%%%%%%%%%%%%%%%%%%%%%%%%%%%%%%%%%%%%%%%%%%%%%%%%%%%%%%%%%%%
\subsection{THE PSEUDOPARTICLE, HOLON, AND SPINON LANDAU-LIQUID
ENERGY FUNCTIONAL}

The energy Landau-liquid functional introduced in Ref.
\cite{Carmelo97} is generated by expanding the $c$ and $\alpha
,\nu$ pseudoparticle band-momentum distribution functions $N_c
(q)$ and $\{N_{\alpha,\,\nu} (q)\}$ where $\alpha =c,\,s$ and
$\nu=1,2,3,...$. Such an expansion is performed around the
ground-state values of these functions given in Appendix C of Ref.
\cite{I} and inserting these distributions in the energy
functional $E$ associated with the Hamiltonian (\ref{H}). Such an
energy functional is given by

\begin{equation}
E=E_H + \sum_{\alpha =c,\,s}\mu_{\alpha}\,S^{\alpha}_z + {U\over
2}[L_{c,\,+1/2}-L_{c,\,-1/2}-{N_a\over 2}] \, ,
\label{EGS}
\end{equation}
where $E_H$ is the functional (\ref{EH}), the parameters
$\mu_{\alpha}$ are the same as in Eq. (\ref{H}), and
$S^{\alpha}_z$ are the eigenvalues of the diagonal generators
defined in Eq. (\ref{Sz}). Let us consider that the initial ground
state corresponds to values of the electronic density such that
$0\leq n\leq 1$ and values of the spin density such that $0\leq
m\leq n$. Throughout this paper we consider such ground states.
The studies of Ref. \cite{Carmelo97} have not expressed the energy
Landau-liquid functional in terms of the deviations in the value
of the $-1/2$ Yang holon and $-1/2$ HL spinon numbers. That
problem was first addressed in Ref. \cite{I}. As discussed in that
reference, the general excitation spectrum of interest for the
problem of the finite-energy correlation functions can be written
as the sum of a finite energy $\omega_0$ and of a gapless
contribution expressed in terms of the pseudoparticle energy
bands. Such a energy Landau-liquid functional reads,

\begin{eqnarray}
\Delta E & = & \omega_0 + {N_a\over 2\pi} \int_{q^{-}_c}^{q^{+}_c}
dq\,\epsilon_c (q)\,\Delta N_c (q) + {N_a\over 2\pi}
\int_{-q_{s,\,1}}^{+q_{s,\,1}} dq\,\epsilon_{s ,\,1}(q)\,\Delta
N_{s,\,1}(q) \nonumber \\ & + & {N_a\over 2\pi}\sum_{\alpha
=c,\,s}\,\sum_{\nu =1+\delta_{\alpha ,\,s}}^{\infty}
\int_{-q_{\alpha,\,\nu}}^{+q_{\alpha,\,\nu}}
dq\,\epsilon^0_{\alpha ,\,\nu}(q)\,\Delta N_{\alpha,\,\nu}(q) \, ,
\label{E1GS}
\end{eqnarray}
where $\Delta N_c (q)$, $\Delta N_{c,\,\nu} (q)$, and $\Delta
N_{s\,\nu} (q)$ are the pseudoparticle band-momentum distribution
function deviations. The band-momentum limiting values
$q_{s,\,1}$, $q_{\alpha,\,\nu}$, and $q_{c}^{\pm}$ are defined in
Eqs. (B14)-(B17) of Ref. \cite{I}, whereas the pseudoparticle
energy bands $\epsilon_c (q)$, $\epsilon_{s ,\,1}(q)$, and
$\epsilon^0_{\alpha ,\,\nu}(q)$ are defined by Eqs. (C15)-(C21) of
the same reference.

The expression for the energy parameter $\omega_0$ on the
right-hand side of Eq. (\ref{E1GS}) which controls the
finite-energy physics is determined by the number of $-1/2$ holons
and number of $-1/2$ spinons (except those associated with $s,1$
pseudoparticles) of the final states relative to the ground state.
Creation of a $-1/2$ holon requires an amount of energy $2\mu$ and
momentum $\pi$ whereas creation of a $-1/2$ spinon (except those
which are part of $s,1$ pseudoparticles) requires an amount of
energy $2\mu_0 H$ and zero momentum. Thus the energy parameter
$\omega_0$ is a simple expression in terms of the deviations in
the numbers of $-1/2$ holons and $-1/2$ spinons which reads,

\begin{equation}
\omega_0 = 2\mu\, \Delta M_{c,\,-1/2} + 2\mu_0\,H\, [\Delta
M_{s,\,-1/2}-\Delta N_{s,\,1}] \, ,
\label{om0}
\end{equation}
where $\Delta M_{\alpha ,\,-1/2}\equiv [M_{\alpha ,\,-1/2} -
M^{0}_{\alpha ,\,-1/2}]$ are the deviations in the numbers of
$-1/2$ holons ($\alpha =c$) and of $-1/2$ spinons ($\alpha =s$)
and $\Delta N_{s,\,1} \equiv [N_{s,\,1}-N^0_{s,\,1}]$ is the
deviation from the ground state value of the number of $s,1$
pseudoparticles. The ground-state pseudoparticle numbers are given
in Appendix C of Ref. \cite{I}.

The final states associated with the energy functional
(\ref{E1GS}) belong to different CPHS ensemble spaces, depending
on the numbers of quantum objects of these states. According to
the form of the general energy spectrum defined in Eqs.
(\ref{E1GS}) and (\ref{om0}), final states belonging to CPHS
ensemble spaces with finite occupancies of $-1/2$ holons have
finite excitation energy relative to the ground state. These
states belong to CPHS ensemble subspaces which are spanned by
states with finite occupancies of $-1/2$ Yang holons and/or of
$c,\nu$ pseudoparticles. Again following expression (\ref{om0}),
for finite values of the spin density $m$, states belonging to
CPHS ensemble subspaces spanned by states with finite occupancies
of $-1/2$ HL spinons and/or of $s,\nu$ pseudoparticles belonging
to $\nu >1$ branches, have also finite excitation energy relative
to the ground state. However, most of the results of the present
paper consider the limit of zero spin density, $m=0$, where the
energy spectrum of these states is gapless.

On the other hand, for general values of $n$ and $m$ the
low-energy Hilbert subspace corresponds to the CPHS ensemble space
which contains the ground state. The corresponding ground-state
$c$ pseudoparticle, $-1/2$ holon, and $-1/2$ spinon numbers are
given in Eqs. (C24) and (C25) of Ref. \cite{I}. Such a CPHS
ensemble space is spanned by states with finite pseudoparticle
occupancy for the $c$ and $s,\,1$ pseudoparticle bands only. We
recall that according to the results of Ref. \cite{I} the $s,1$
pseudoparticle is a composite quantum object. It is constituted of
two spinons of opposite spin projection. In the studies of Refs.
\cite{Carmelo91,Carmelo92,Carmelo99} the $c$ and $s,1$
pseudoparticle branches were referred to as $c$ pseudoparticles
and $s$ pseudoparticles respectively. The excited states contained
in that CPHS-ensemble space have band-momentum distribution
functions whose occupancies are different from the ones of the $c$
and $s,1$ ground-state distributions provided in Eqs. (C1) and
(C2) respectively, of Ref. \cite{I}. At fixed spin $\sigma$
electron numbers, the occupancies of these excited states can be
generated from the ground state by pseudoparticle - pseudoparticle
hole processes in the $c$ and $s,1$ bands.

%%%%%%%%%%%%%%%%%%%%%%%%%%%%%%%%%%%%%%%%%%%%%%%%%%%%%%%%%%%%%%%%%%%%%%%%%%
\subsection{THE GROUND-STATE ELECTRON DOUBLE OCCUPATION}

In the particular case of a zero spin-density ground state, from
the use of the general Eq. (\ref{Dee}) we find

\begin{equation}
D_0 = {\partial E_{GSH}\over\partial U}  = {N\over
2}\,\Bigl({n\over 2}\Bigr)\,f(n,U/t) \, ,
\label{D0}
\end{equation}
where $D_0$ denotes the ground-state electron double occupation
and $E_{GSH}$ stands for the ground-state energy,

\begin{equation}
E_{GSH} = -2t {N_a\over 2\pi} \int_{q_{Fc}^{-}}^{q_{Fc}^{+}} dq\,
\cos k^{0}(q) = -2t {N_a\over 2\pi} \int_{-Q}^{Q} dk\,2\pi\rho_c
(k)\, \cos k  \, .
\label{EGSH}
\end{equation}
Here the Fermi band-momenta $q_{Fc}^{\pm}$ are defined by Eqs.
(C4)-(C7) of Ref. \cite{I} and the ground-state functions
$k^{0}(q)$ and $2\pi\rho_c (k)$ are studied below. This energy
expression corresponds to the Hamiltonian (\ref{HH}) and is
obtained by use of the ground-state numbers and deviations of Eqs.
(C1)-(C3) and (C24)-(C25) of Ref. \cite{I} in the general energy
functional (\ref{EH}). In order to study the ground-state double
occupation (\ref{D0}), let us provide useful information on the
related energy (\ref{EGSH}). The ground-state function $k^{0}(q)$
on the right-hand side of Eq. (\ref{EGSH}) is related to the
distribution $2\pi\rho_c (k)$ considered below by the following
equation,

\begin{equation}
{\partial k^{0}(q)\over \partial q} = {1\over 2\pi\rho_c
\Bigl(k^{0}(q)\Bigr)} \, .
\label{rhoc}
\end{equation}
The function $k^{0}(q)$ is also related to the ground-state
rapidity functions $\Lambda^{0}_{\alpha,\,\nu}(q)$ such that,

\begin{equation}
{\partial \Lambda^{0}_{\alpha,\,\nu}(q)\over\partial q} = {1\over
2\pi\sigma_{\alpha,\,\nu}\Bigl(\Lambda^{0}_{\alpha,\,\nu}(q)\Bigr)}
\, ; \hspace{1.0cm} \alpha = c,s \, , \hspace{0.5cm} \nu =
1,2,3,... .
\label{sigan}
\end{equation}
The above functions $2\pi\rho_c(k)$,
$2\pi\sigma_{c,\,\nu}(\Lambda)$, and
$2\pi\sigma_{s,\,\nu}(\Lambda)$ are the solutions of the following
integral equations,

\begin{equation}
2\pi\rho_c(k) = 1 + {4t\,\cos k\over \pi} \int_{-B}^{B}d\Lambda\,
{U \over U^2+(4t)^2(\Lambda - \sin
k)^2}\,2\pi\sigma_{s,\,1}(\Lambda) \, ,
\label{rhocex}
\end{equation}

\begin{equation}
2\pi\sigma_{c,\,\nu}(\Lambda) = 2 Re \Bigl({4t\over \sqrt{(4t)^2 -
(4t\,\Lambda - i\nu U)^2}}\Bigl) - {4t\over \pi}\int_{-Q}^{Q}dk\,
{\nu U\over (\nu U)^2+(4t)^2(\sin k -\Lambda)^2}\,2\pi\rho_c(k) \,
,
\label{sigcnex}
\end{equation}
and

\begin{eqnarray}
2\pi\sigma_{s,\nu}(\Lambda) & = & {4t\over \pi} \int_{-Q}^{Q}dk\,
{\nu U\over (\nu U)^2+(4t)^2(\sin k -\Lambda)^2}\,2\pi\rho_c (k)
\nonumber \\
& - & {2t\over \pi U}
\int_{-B}^{B}d\Lambda'\,\Theta^{[1]}_{1,\,\nu } \Bigl({4t(\Lambda
-\Lambda')\over U}\Bigl)\,2\pi\sigma_{s,\,1}(\Lambda') \, ,
\label{sigsnex}
\end{eqnarray}
where the expression of the function $\Theta^{[1]}_{1,\,\nu }(x)$
and the parameters $Q$ and $B$ are given in Eqs. (C22) and (C23)
of Ref. \cite{I}. We note that $2\pi\rho_c(k)$ and
$2\pi\sigma_{s,\,1}(\Lambda)$ equal the functions $2\pi\rho (k)$
and $2\pi\sigma (\Lambda)$ respectively, introduced by Lieb and Wu
\cite{Lieb}. At half filling and spin density $m=0$ the parameters
$Q$ and $B$ read $Q=\pi$ and $B=\infty$ and Eqs. (\ref{rhocex})
-(\ref{sigsnex}) can be solved by Fourier transform. Use of the
obtained expressions in Eq. (\ref{rhoc}) and in Eq. (\ref{sigan})
for $\alpha ,\nu =s, 1$ leads to the following expressions,

\begin{equation}
q = k^0 (q) + 2\int_{0}^{\infty}dx\,J_0(x)\,{\sin \Bigl(x\,\sin
k^0 (q)\Bigr)\over x\,[1+e^{xU \over 2t}]} \, ; \hspace{0.5cm}
\vert q\vert \leq \pi  \, ,
\label{k0n1}
\end{equation}
and

\begin{equation}
q = \int_{0}^{\infty}dx\, J_0(x)\,{\sin \Bigl(x\,\Lambda^0_{s,\,1}
(q)\Bigr) \over x\,\cosh({xU\over 4t})} \, ; \hspace{0.5cm} \vert
q\vert \leq \pi/2 \, ,
\label{Ls1n1}
\end{equation}
where $J_0 (x)$ and $J_1 (x)$ are Bessel functions. These
expressions define the inverse functions of $k^0 (q)$ and
$\Lambda^0_{s,\,1} (q)$ respectively. Based on these expressions
it is straightforward to find that at half filling and zero spin
density the energy (\ref{EGSH}) can be written in closed form with
the result \cite{Lieb},

\begin{equation}
E_{GSH} = -4N_a\,t\int_{0}^{\infty} dx\, {J_0 (x)\,J_1 (x)\over
x[1 + e^{x U\over 2t}]} \, .
\label{EGSHn1}
\end{equation}

One can use the obtained ground-state energy expressions in the
evaluation of the electron ground-state double occupation
(\ref{D0}). We find that in the limits of vanishing and infinite
on-site repulsion $U/t$ the function $f(n,U/t)$ of Eq. (\ref{D0})
is given by $1$ and $0$ respectively. Closed form expressions for
that function can be obtained for electronic densities $0\leq
n\leq 1$ in the limits $U/t\rightarrow 0$ and $U/t>> 1$, with the
result,

\begin{eqnarray}
f (n,U/t) & = & 1 \, ;
\hspace{4.4cm} U/t\rightarrow 0 \, , \nonumber \\
f (n,U/t) & = & \Bigl({4t\over U}\Bigr)^2\,\ln
2\,\Bigl(1-{\sin\Bigl(2\pi n\Bigr)\over 2\pi n}\Bigr) \, ;
\hspace{1cm} U/t >> 1 \, .
\label{flim}
\end{eqnarray}
For the specific case of half filling, $n=1$, one finds from the
use of Eqs. (\ref{D0}) and (\ref{EGSHn1}) the following
closed-form expression for the function $f(n,U/t)$
\cite{Carmelo88},

\begin{equation}
f (1,U/t) = 4 \int_{0}^{\infty} dx\, {J_0 (x)\,J_1 (x)\over 1 +
\cosh ({x U\over 2t})} \, .
\label{fn1}
\end{equation}

According to the general expression (\ref{Tee}), the ground-state
kinetic energy $T_0$ is given by $T_0 = t\,\partial
E_{GSH}/\partial t$ \cite{Carmelo86}, where $E_{GSH}$ stands again
for the ground-state energy associated with the Hamiltonian
(\ref{HH}).

%%%%%%%%%%%%%%%%%%%%%%%%%%%%%%%%%%%%%%%%%%%%%%%%%%%%%%%%%%%%%%%%
\subsection{THE ELECTRON DOUBLE OCCUPATION FUNCTIONAL}

Our goal is the study of the deviations $\Delta D$ from the
ground-state electron double occupation $D_0$ given in Eq.
(\ref{D0}), generated by changes in the numbers of
pseudoparticles, $-1/2$ Yang holons, and $-1/2$ HL spinons
associated with transitions to other energy eigenstates. Such a
study provides information on the $U/t$ dependence of the electron
site distribution configurations of the real-space lattice which
describe the energy eigenstates. The above-mentioned deviations
are interesting quantities which provide useful information about
the localization/delocalization degree of the corresponding
elementary quantum objects. Fortunately, the electron
double-occupation $D$ of the final energy eigenstates can be
obtained by means of the general expression (\ref{Dee}). Thus one
can compute the electron double occupation deviations $\Delta
D\equiv [D-D_0]$. These can be obtained simply by taking the $U$
derivative of the excitation energy defined by the following
Landau-liquid energy functional,

\begin{equation}
\Delta E_H = \Delta E - \sum_{\alpha =c,\,s}\mu_{\alpha}\,\Delta
S^{\alpha}_z - {U\over 2}[\Delta L_{c,\,+1/2}-\Delta L_{c,\,-1/2}]
\, ,
\label{EF}
\end{equation}
where $\Delta E$ is the energy functional (\ref{E1GS}), $\Delta
L_{c,\,\pm 1/2}$ denotes the deviations in the numbers of $\pm
1/2$ Yang holons, and the remaining quantities are the same as on
the right-hand side of Eq. (\ref{EH}).

The deviations $\Delta D$ can be written in functional form. To
first order in the $-1/2$ Yang holon and $-1/2$ HL spinon number
deviations and pseudoparticle band-momentum distribution function
deviations, the double-occupation functional reads,

\begin{eqnarray}
\Delta D & = & \sum_{\alpha =c,s}\,\Delta
L_{\alpha,\,-1/2}\,D_{\alpha\,,-1/2} + {N_a\over
2\pi}\int_{q_c^{-}}^{q_c^{+}}\,dq\, \Delta N_c (q)\, D_c (q)
\nonumber \\ & + & {N_a\over 2\pi}\sum_{\alpha =c,\,s}\,\sum_{\nu
=1}^{\infty}\int_{-q_{\alpha,\,\nu }}^{q_{\alpha,\,\nu }}\,dq\,
\Delta N_{\alpha,\,\nu}(q) D_{\alpha,\,\nu} (q)\, \, .
\label{DD}
\end{eqnarray}
Here $\Delta L_{\alpha,\,-1/2}$ is the deviation from the
ground-state number of $-1/2$ Yang holons ($\alpha =c$) and $-1/2$
HL spinons ($\alpha =s$) and $\Delta N_c (q)$ and $\Delta
N_{\alpha,\,\nu}(q)$ are the $c$ and $\alpha ,\nu$ pseudoparticle
band-momentum distribution function deviations given in Eq. (58)
of Ref. \cite{I} respectively. On the right-hand side of Eq.
(\ref{DD}) $D_{\alpha\,,-1/2}$ is the $-1/2$ Yang holon ($\alpha
=c$) and $-1/2$ HL spinon ($\alpha =s$) double occupation, while
$D_c (q)$ and $D_{\alpha,\,\nu}(q)$ denote the $c$ pseudoparticle
and $\alpha ,\nu$ pseudoparticle double occupation spectra
respectively. These double occupations and spectra equal the
corresponding deviation in the value of double occupation
(\ref{D}) which results from creation of a $-1/2$ Yang holon or
$-1/2$ HL spinon and of creation of a $c$ or $\alpha ,\nu$
pseudoparticle at band-momentum $q$ respectively. The
pseudoparticle double-occupation spectra can be expressed as,

\begin{equation}
D_c (q) = {\partial\,{\bar{\epsilon }}_c (q)\over\partial\,U}  \,
; \hspace{1cm} D_{\alpha,\,\nu}(q) = {\partial\,{\bar{\epsilon
}}_{\alpha,\,\nu} (q)\over\partial\,U}  \, ,
\label{DcDan}
\end{equation}
where the pseudoparticle bands ${\bar{\epsilon }}_c (q)$ and
${\bar{\epsilon }}_{\alpha,\,\nu} (q)$ are given by,

\begin{eqnarray}
{\bar{\epsilon }}_c (q) & = & \epsilon_c (q) + U/2 - \mu + \mu_0\,
H\, ; \hspace{1cm} {\bar{\epsilon }}_{s,\,1} (q) =
\epsilon_{s,\,1}
(q) - 2\mu_0\,H \, ; \nonumber \\
{\bar{\epsilon }}_{c,\,\nu} (q) & = & \epsilon_{c,\,\nu}^0 (q) +
\nu U \, ; \hspace{1cm} {\bar{\epsilon }}_{s,\,\nu} =
\epsilon_{s,\,\nu}^0 (q) \, ; \hspace{0.5cm} \nu >1 \, .
\label{barepsi}
\end{eqnarray}
Here the pseudoparticle energy bands $\epsilon_c (q)$,
$\epsilon_{s,\,1} (q)$, $\epsilon_{c,\,\nu}^0 (q)$, and
$\epsilon_{s,\,\nu}^0 (q)$ for $\nu >1$ are defined by Eqs.
(C15)-(C21) of Ref. \cite{I}. These are the energy bands of the
energy functional (\ref{E1GS}). The $-1/2$ Yang holon and $-1/2$
HL spinon electron double-occupation deviation numbers are simply
given by

\begin{equation}
D_{c\,,-1/2} = 1 \, ; \hspace{1cm} D_{s\,,-1/2} = 0  \, .
\label{Dhs}
\end{equation}

As for electron double occupation, we could also introduce a
kinetic-energy functional associated with the kinetic-energy
deviations $\Delta T\equiv [T-T_0]$. However, the only information
we need for our present study is whether creation of an elementary
quantum object leads to a finite or a vanishing kinetic-energy
deviation $\Delta T$.

The pseudoparticle electron double occupation spectra
(\ref{DcDan}) and $-1/2$ Yang holon and $-1/2$ HL spinon electron
double occupation (\ref{Dhs}) provide interesting physical
information about the degree of localization/delocalization of the
corresponding quantum objects. For instance, Eq. (\ref{Dhs})
reveals that for the whole parameter space creation of a $-1/2$
Yang holon leads to $\Delta D=1$. That excitation involves
creation of two electrons. From the expression of the $\eta$-spin
flip generators given in Eq. (\ref{Sc}), it follows that creation
of a $-1/2$ Yang holon involves the creation of an extra on-site
electron pair on a ground-state empty site for all real-space
lattice electron site distribution configurations which describe
the initial state. Consistently, we find that creation of a $-1/2$
Yang holon leads to a kinetic-energy deviation $\Delta T=0$. Since
creation of such a quantum objects leads to no change in the value
of the ground-state kinetic energy, we say that the $-1/2$ Yang
holon carries no kinetic energy. Therefore, in spite of having
charge $-2e$, $-1/2$ Yang holons do not contribute directly to
charge transport, as mentioned in Ref. \cite{I}. This concept of
an elementary quantum object {\it carrying} kinetic energy when
its creation leads to a finite deviation $\Delta T$, is inspired
in the electronic conductivity sum rule. That sum rule refers to
the frequency dependent electronic conductivity and states that it
is proportional to $\vert T\vert$ \cite{Carmelo86}. Thus if
creation of a quantum object does not change the value of $T$, it
follows that the conductivity sum rule remains unchanged.
Therefore, we say that such a quantum object {\it carries no}
kinetic energy. Furthermore, we find from analysis of Eq.
(\ref{Dhs}) that creation of a $-1/2$ HL spinon leads to no change
in the ground-state double occupation. This is consistent with the
form of the generators defined in Eq. (\ref{Ss}). Thus the spin
flip associated with creation of a $-1/2$ HL spinon corresponds to
electrons located at singly occupied sites for all real-space
lattice electron site distribution configurations which describe
the initial ground state.

Let us start the discussion of the pseudoparticle electron
double-occupation spectra defined in Eq. (\ref{DcDan}) by
considering the limit $U/t\rightarrow\infty$. These spectra can be
easily evaluated and in that limit the elementary quantum objects
have simpler expressions in terms of real-space lattice electron
site distribution configurations. As discussed in Refs.
\cite{I,III}, such $U/t\rightarrow\infty$ configurations are the
same which describe the energy eigenstates for all values of $U/t$
in terms of an effective electronic lattice. Such a lattice refers
to the occupancy configurations of the rotated electrons
introduced in Ref. \cite{III}. It has the same number of sites
$N_a$ and lattice constant $a$ as the real-space lattice
\cite{I,III}. In the limit of $U/t\rightarrow\infty$ electron
double occupation $D$ is a good quantum number which equals the
number of rotated-electron doubly-occupied sites \cite{I,III}. We
recall that for rotated electrons double occupation is a good
quantum number for all values of $U/t$. As a result we find that
as $U/t\rightarrow\infty$ the electron double-occupation
functional (\ref{DD}) simply reads,

\begin{equation}
\Delta D = \Delta M_{c,\,-1/2} = \Delta L_{c,\,-1/2}+ {N_a\over
2\pi}\sum_{\nu =1}^{\infty}\int_{-q_{c,\,\nu }}^{q_{c,\,\nu
}}\,dq\,\nu\,\Delta N_{c,\,\nu} (q) \, ; \hspace{1cm}
U/t\rightarrow\infty \, .
\label{DDUi}
\end{equation}
From comparision of Eqs. (\ref{DD}) and (\ref{DDUi}), we arrive to
the following $U/t\rightarrow\infty$ expressions for the
pseudoparticle electron double-occupation spectra,

\begin{equation}
D_c (q) = D_{s,\,\nu}(q) = 0 \, ; \hspace{1cm} D_{c,\,\nu} (q) =
\nu \, ; \hspace{1cm} U/t\rightarrow\infty \, .
\label{Dcaq}
\end{equation}
These electron double-occupation spectra expressions show that
increasing the number of $c$ pseudoparticles by one does not
change electron double occupation when $U/t\rightarrow\infty$. On
the other hand, from Eq. (\ref{Dcaq}) we find that increasing the
number of $c,\nu$ pseudoparticles by one leads to an increase
$\nu$ in electron double occupation $D$. Moreover, we also studied
the corresponding deviations in kinetic energy and found that in
this limit the $c$ pseudoparticles ($c,\nu$ pseudoparticles) carry
kinetic energy (carry no kinetic energy).

These results reveal that in the present limit, the $c$
pseudoparticles correspond to electron singly occupied sites and
that the $-1/2$ holons and $+1/2$ holons correspond to electron
pairs and electronic hole pairs at the same real-space lattice
site respectively. In reference \cite{III} it is found that such a
property is closely related to the rotated-electron effective
electronic lattice mentioned above. As we discuss below, we find
that at finite values of $U/t$ a composite $2\nu$-holon $c,\nu$
pseudoparticle with band-momentum $q$ such that $\vert q\vert<[\pi
-2k_F]$ carries kinetic energy. This means that for finite values
of $U/t$ the $2\nu$ extra electrons are not always created at the
same site and thus that composite quantum object contributes to
charge transport. However, we find that $\Delta T$ is a decreasing
function of $U/t$ which vanishes in the limit
$U/t\rightarrow\infty$. Thus creation of a $c,\nu$ pseudoparticle
leads in that limit to $\Delta T=0$ and these objects become
localized and carry no kinetic energy. It follows that in spite of
having charge $-2\nu e$, for $U/t\rightarrow\infty$ the $c,\nu$
pseudoparticles do not contribute directly to charge transport, as
mentioned in Ref. \cite{I}. In that limit they describe a number
$\nu$ of localized electron pairs, each pair localized at the same
site.

On the other hand, in agreement with the results of Ref. \cite{I}
the chargeons have charge $-e$ and we find that in the
$U/t\rightarrow\infty$ limit they carry kinetic energy and thus
are the only carriers of charge. When $U/t\rightarrow\infty$ the
finite-energy Hilbert subspace is spanned by states with $D=0$
because it requires an infinite energy, $D\,U$, to excite the
system into states with $D>0$ electron double occupation. In this
limit, the $c$ pseudoparticles become the spinless fermions of
Refs. \cite{Ogata,Ricardo,Penc95,Penc96}. The associated chargeons
describe the charge excitations associated with the electron
singly occupied sites of the real-space lattice. The spin degrees
of freedom of the electron singly occupied sites are described by
the spinons. In the $U/t\rightarrow\infty$ limit creation of $\pm
1/2$ spinons describes spin flips of localized electrons at singly
occupied sites. Therefore, creation of spinons does not contribute
to double occupation, as confirmed by the results of Eq.
(\ref{DDUi}) for the $2\nu$-spinon composite $s,\nu$
pseudoparticles. This holds true for all finite values of $U/t$ in
the case of the HL spinons, as confirmed by Eq. (\ref{Dhs}).

%%%%%%%%%%%%%%%%%%%%%%%%%%%%%%%%%%%%%%%%%%%%%%%%%%%%%%%%%%%%%%%%
\subsection{ELECTRON DOUBLE OCCUPATION SPECTRA FOR FINITE VALUES OF U/t AND ZERO SPIN DENSITY}

Let us now study the pseudoparticle electron double-occupation
spectra of Eq. (\ref{DcDan}) for finite values of $U/t$. For
simplicity, let us again consider the case of zero-spin density
$m=0$. In that case we find the following relations, which are
valid for values of electronic density in the domain $0\leq n\leq
1$ and for all values of $U/t$,

\begin{equation}
D_{s,\,\nu} (q) = 0 \, , \hspace{0.5cm} \nu > 1 \, ,
\label{Deq0}
\end{equation}

\begin{eqnarray}
D_{c,\,\nu} (0) & \leq & D_{c,\,\nu} (q) \leq D_{c,\,\nu} (\pm
[\pi -2k_F]) = \nu \, ; \nonumber \\
D_{s,\,1} (\pm k_F) = 0 & \leq & D_{s,\,1} (q) \leq D_{s,\,1} (0)
\, .
\label{ineqD}
\end{eqnarray}

The electron double-occupation spectra are derived from the
expressions of Eq. (\ref{barepsi}) for the energy bands
${\bar{\epsilon }}_c (q)$ and ${\bar{\epsilon}}_{c,\,\nu}(q)$ by
means of the relation provided in Eq. (\ref{DcDan}). Therefore,
the features of these bands fully control and determine the
band-momentum $q$, electronic density $n$, and on-site repulsion
$U/t$ dependence of the pseudoparticle electron double-occupation
spectra. As we discuss in the ensuing section, the energy width of
the band ${\bar{\epsilon}}_c (q)$ is independent of $U/t$, being
constant for all values of $U/t$ and $n$ and is given by
$W_c\equiv\vert{\bar{\epsilon}}_c (\pi)-{\bar{\epsilon}}_c
(0)\vert=4t$. In contrast, the width
$W_{c,\,\nu}\equiv\vert{\bar{\epsilon}}_{c,\,\nu} (\pi
-2k_F)-{\bar{\epsilon}}_{c,\,\nu} (0)\vert$ of the $c,\nu$
pseudoparticle energy bands is a decreasing function of $U/t$,
vanishing as $U/t\rightarrow\infty$. The same behavior occurs also
for the $s,1$ pseudoparticle band, which is such that
$W_{s,\,1}\equiv\vert{\bar{\epsilon}}_{s,\,1}
(k_F)-{\bar{\epsilon}}_{s,\,1} (0)\vert\rightarrow 0$ as
$U/t\rightarrow\infty$.

This qualitative difference between the $c$ pseudoparticle and the
remaining pseudoparticle energy bands leads to qualitative
differences in the corresponding electron double-occupation
spectra as well. For instance, there are qualitative differences
between the $q$ dependence of the electron double-occupation
spectrum $D_c (q)$ on the one hand, and that of the electron
double-occupation spectra $D_{s,\,1}(q)$ and $D_{c,\,\nu}(q)$ on
the other hand. While the function $D_c (q)$ changes slowly with
band-momentum $q$, for finite values of $U/t$ the functions
$D_{s,\,1}(q)$ and $D_{c,\,\nu}(q)$ show a significant $q$
dependence. For instance, the values $D_{s,\,1} (k_F)=0$ and
$D_{c,\,\nu} (\pi-2k_F)=\nu$ are $U/t$ independent, whereas the
values of $D_{s,\,1} (0)$ and $D_{c,\,\nu} (0)$ depend on this
quantity. In contrast, the $U/t$ dependence of the spectrum $D_c
(q)$ is very similar for all band-momentum values $q$. For finite
values of $U/t$ the electron double-occupation spectrum $D_{s,\,1}
(q)=D_{s,\,1} (-q)$ is a decreasing function of $\vert q\vert$,
having its minimum value $D_{s,\,1} (k_F)=0$, at $\vert
q\vert=q_{s,\,1}=k_F$ and its maximum value at $\vert q\vert =0$.
For finite values of $U/t$ the electron double-occupation spectrum
$D_{c,\,\nu} (q)=D_{c,\,\nu} (-q)$ is an increasing function of
$\vert q\vert$, having its minimum value at $\vert q\vert=0$ and
its maximum value $D_{c,\,\nu} (\pi-2k_F)=\nu$, at $\vert q\vert =
q_{c,\,\nu}= [\pi -2k_F]$.

Since both the electron double occupation spectra $D_{s,\,1}(q)$
and $D_{c,\,\nu} (q)$ are not dependent on $U/t$ for the limiting
band-momentum values $q=\pm k_F$ and $q=\pm [\pi - 2k_F]$
respectively, and show the strongest $U/t$ dependence for $q=0$,
below we consider mainly the $U/t$ and $n$ dependences of the
$q=0$ parameters $D_{s,\,1}(0)$ and $D_{c,\,\nu} (0)$. In
agreement with the inequalities of Eq. (\ref{ineqD}), for
increasing values of the band-momentum absolute value $\vert
q\vert$, these spectra reach smoothly the $U/t$-independent
limiting values $D_{s,\,1} (\pm k_F) = 0$ and $D_{c,\,\nu} (\pm
[\pi -2k_F]) = \nu$ respectively. According to the results
obtained in Ref. \cite{III}, at $m=0$ only the $c$ pseudoparticle,
$s,1$ pseudoparticle, $c,1$ pseudoparticle, and $c,2$
pseudoparticle bands contribute significantly to the energy
spectra of one-electron and two-electron excitations. Thus, below
we limit our studies of the spectrum $D_{c,\,\nu} (0)$ to the
branches $\nu =1$ and $\nu =2$. In the case of the $D_c (q)$
spectrum, we consider the $n$ and $U/t$ dependence of both the
parameters $D_c (0)$ and $D_c (-2k_F)=D_c (2k_F)$. Although the
spectrum $D_c (q)$ is gently dependent on $q$, in addition to the
value at the band-momentum $q=0$ we also consider the value at
$\pm 2k_F$. On the one hand, we study the parameter $D_c (2k_F)$
because creation and annihilation of $c$ pseudoparticles at the
Fermi points $\pm 2k_F$ are associated with important one-electron
and two-electron excitations \cite{optical}. On the other hand,
since for the other pseudoparticle branches we study the $U/t$ and
$n$ dependence of the electron double-occupation spectrum at
$q=0$, for completeness we also consider in our study the
parameter $D_c (0)$.

In figures 1 and 2 the electron double-occupation pseudoparticle
parameters $D_c (0)$ and $D_c (2k_F)$ are plotted as a function of
$U/t$ for different values of the electronic density $n$. The
electron double-occupation pseudoparticle parameters
$D_{s,\,1}(0)$, $D_{c,\,1}(0)$, and $D_{c,\,2}(0)$ are also
plotted as a function of $U/t$ and for the same values of the
electronic density in Figs. 3-5. The curves plotted in these
figures show that $D_c (0)$ and $D_c (2k_F)$ are decreasing
functions of $U/t$. Both of these functions have the maximum value
$D_c (0)=D_c (2k_F)=n/2$, for $U/t\rightarrow 0$ and vanish in the
limit $U/t\rightarrow\infty$. On the other hand, $D_{s,\,1}(0)$
vanishes both for $U/t\rightarrow 0$ and $U/t\rightarrow\infty$
and has its $n$ dependent maximum value for a value of $U/t$
between $1$ and $4$. While for small values of the electronic
density $n$ the parameters $D_{c,\,1}(0)$ and $D_{c,\,2}(0)$ are
decreasing functions of $U/t$, for values of $n$ close to $1$
there is a minimum value at a small but finite value of $U/t$.
These parameters are given by $n$ and $2n$ respectively for
$U/t\rightarrow 0$, and tend to $1$ and $2$ respectively as
$U/t\rightarrow\infty$.

Before discussing the physical information contained in the curves
plotted in Figs. 1-5, we consider some limiting cases where we
could evaluate analytical expressions for the pseudoparticle
electron double-occupation spectra. In addition, before that
discussion we also indicate the cases where creation of the
elementary quantum objects leads to a vanishing kinetic-energy
deviation $\Delta T$. In the limit $U/t\rightarrow 0$ we find the
following expressions valid for densities $0\leq n\leq 1$,

\begin{equation}
D_c (0) = D_c (\pm 2k_F) = n/2 \, ; \hspace{1cm} D_{s,\,1} (0) = 0
\, ; \hspace{1cm} D_{c,\,\nu}(0) = \nu\,n \, .
\label{DU0}
\end{equation}

On the other hand, in the limit $U/t >> 1$ the same quantities can
be written as follows,

\begin{eqnarray}
D_c (0) & = & \Bigl({2t\over U}\Bigr)^2\,n\,\ln (2)\,\Bigl[1-{\sin
(2\pi n)\over 2\pi n}\Bigr] \, ; \nonumber \\ D_c (\pm 2k_F) & = &
\Bigl({2t\over U}\Bigr)^2\,n\,\ln (2)\,\Bigl[2\sin^2 (\pi n) + 1 -
{\sin (2\pi n)\over 2\pi n}\Bigr] \, ; \nonumber \\ D_{s,\,1} (0)
& = & \Bigl({2t\over U}\Bigr)^2\,{\pi n\over 2}\,\Bigl[1-{\sin
(2\pi n)\over 2\pi n}\Bigr]\, ; \nonumber \\ D_{c,\,\nu} (0) & = &
\nu - \Bigl({4t\over U}\Bigr)^2\,{1\over 2\nu}\,\Bigl[(1-n) -
{\sin (2\pi(1-n))\over 2\pi}\Bigr] \, .
\label{DlU}
\end{eqnarray}

Moreover, the following expressions are valid for all values of
$U/t>0$ in the limit of vanishing density $n\rightarrow 0$,

\begin{eqnarray}
D_c (0) & = & D_c (\pm 2k_F) = D_{s,\,1} (0) = 0 \, ; \nonumber \\
D_{c,\,\nu} (0) & = & \nu\,{\nu\,U\over \sqrt{(4t)^2 +(\nu\,U)^2}}
\, . \label{Dn0}
\end{eqnarray}

In the case of density $n=1$ we could derive the following
expressions, which are valid for all values of $U/t>0$,

\begin{eqnarray}
D_c (0) & = & D_c (\pm 2k_F) = \int_{0}^{\infty }dx\,{J_1(x)\over
1+\cosh ({xU\over 2t})} \, ; \nonumber \\ D_{s,\,1} (0) & = &
{1\over 2}\int_{0}^{\infty } dx\, J_1(x){\sinh ({xU\over 4t})\over
[\cosh({xU\over 4t})]^2} \, ; \hspace{2cm} D_{c,\,\nu} (0) = \nu
\, .
\label{Dn1}
\end{eqnarray}
Comparison of the expressions (\ref{DU0})-(\ref{Dn1}) with the
curves of the Figs. 1-5, reveals there is agreement between these
expressions and the curves.

As we mentioned above, we can also evaluate the kinetic energy
deviations $\Delta T\equiv [T-T_0]$ associated with elementary
excitations. In the present study we are mostly interested in
whether the deviations $\Delta T$ which result from creation of a
pseudoparticle, $-1/2$ Yang holon, or $-1/2$ HL spinon are finite
or vanish. Importantly, by combining our analysis of the $\Delta
T$ deviation expression with the related properties of the
electron double-occupation spectra, we find that when such a
deviation vanishes, creation of these quantum objects is either
associated with creation of localized electrons or involves
on-site spin-flip processes. From our study of the kinetic energy
(\ref{T}) we find that the deviation $\Delta T$ which results from
creation of pseudoparticles, $-1/2$ Yang holons, or $-1/2$ HL
spinons vanishes in the following cases:

(i) Creation of one or a vanishing densitiy of $s,1$
pseudoparticles at band momentum values $q=\pm k_{F\uparrow}$ for
all values of $U/t$, $n$, and $m$.

(ii) Creation of one or a vanishing densitiy of $c,\nu$
pseudoparticles at band momentum values $q=\pm [\pi - 2 k_F]$ for
all values of $U/t$, $n$, and $m$.

(iii) Creation of one or a vanishing densitiy of $s,\nu$
pseudoparticles belonging to branches such that $\nu >1$ at band
momentum values $q=\pm [k_{F\uparrow}-k_{F\downarrow}]$ for all
values of $U/t$, $n$, and $m$.

(iv) Creation of $-1/2$ Yang holons for all values of $U/t$, $n$,
and $m$.

(v) Creation of $-1/2$ HL spinons for all values of $U/t$, $n$,
and $m$.

(vi) Creation of $\alpha ,\nu$ pseudoparticles with $\alpha =c,s$
and $\nu =1,2,3,...$ for $m=0$, all band-momentum values $q$, all
densities $n$, and $U/t\rightarrow\infty$.

%%%%%%%%%%%%%%%%%%%%%%%%%%%%%%%%%%%%%%%%%%%%%%%%%%%%%%%%%%%%%%%%
\subsection{ELECTRON DOUBLE OCCUPATION SPECTRA AND THE PSEUDOPARTICLE
DEGREE OF LOCALIZATION/DELOCALIZATION}

We close this section with the analysis and a discussion of the
information provided by the dependence on the electronic density
$n$, on-site electronic repulsion $U/t$, and band-momentum $q$ of
the electron double occupation quantities defined in Eqs.
(\ref{ineqD})-(\ref{Dn1}) and Figs. 1-5. We find that the
dependence of the electron double-occupation quantities and
spectra on these parameters provides insight into the degree of
localization/delocalization of the $-1/2$ Yang holons, $-1/2$ HL
spinons, and pseudoparticles.

While only spin $\sigma$ electrons can be created into the
many-electron system or annihilated from it, creation of spin
$\sigma$ electrons generates back-flow effects which can change
the numbers and the occupancy configurations of pseudoparticles,
holons, and spinons \cite{III}. In the absence of electron
addition or removal these elementary quantum objects are confined
within the many-electron system, their occupancy configurations
describing the exact energy eigenstates. As a result of the
non-perturbative character of the electronic correlations, the
description of the pseudoparticles, holons, and spinons in terms
of electron site distribution configurations of the real-space
lattice is a very complex many-body problem \cite{I}.

For finite values of $U/t$ the electron double occupation $D$
defined in Eq. (\ref{D}) is not a good quantum number. This just
means that the ground state and the other energy eigenstates can
be expressed as a superposition of real-space lattice electron
site distribution configurations with different numbers of
electron doubly occupied sites. On the other hand, in the limit
$U/t\rightarrow\infty$ both electron double occupation $D$, double
unoccupation, spin-down single occupation, and spin-up single
occupation become good quantum numbers \cite{I,III}. Thus in this
limit all energy eigenstates can be expressed as a superposition
of real-space lattice electron site distribution configurations
with the same value of electron double occupation $D$. For
instance, in the particular case of the ground state and
finite-energy excited states that value is $D=0$. Thus, the ground
state electron double occupation $D_0$ is such that
$D_0\rightarrow 0$ as $U/t\rightarrow\infty$, as confirmed by the
form of the expression defined in Eqs. (\ref{D0}) and
(\ref{flim}).

It is useful for our discussion to express the deviations $\Delta
N_{\uparrow}$ and $\Delta N_{\downarrow}$ in the electron numbers
in terms of the deviations in the $c$ pseudoparticle, $\alpha
,\nu$ pseudoparticle, $-1/2$ Yang holon, and $-1/2$ HL spinon
numbers. Based on the form of the expressions given in Eqs.
(\ref{NupNM})-(\ref{MLNcs}), we find the following relations,

\begin{equation}
\Delta N_{\uparrow} = \Delta N_c - \Delta L_{s,\,-1/2} + \Delta
L_{c,\,-1/2} + \sum_{\nu =1}^{\infty}\nu\,\Delta N_{c,\,\nu} -
\sum_{\nu =1}^{\infty} \nu\, \Delta N_{s,\,\nu} \, ,
\label{DNu}
\end{equation}
and

\begin{equation}
\Delta N_{\downarrow} = \Delta L_{s,\,-1/2} + \Delta L_{c,\,-1/2}
+ \sum_{\nu =1}^{\infty}\nu\,\Delta N_{c,\,\nu} + \sum_{\nu
=1}^{\infty} \nu\, \Delta N_{s,\,\nu} \, .
\label{DNd}
\end{equation}
We recall that the numbers of $+1/2$ Yang holons and of $+1/2$ HL
spinons are dependent on the numbers of the remaining quantum
objects. In particular, it follows from Eq. (\ref{ScsNN}) that the
deviations $\Delta L_{c,\,+1/2}$ and $\Delta L_{s,\,+1/2}$ can be
expressed as follows,

\begin{eqnarray}
\Delta L_{c,\,+1/2} & = & - \Delta N_c - 2\sum_{\nu =1}^{\infty}
\nu\,\Delta N_{c,\,\nu} - \Delta L_{c,\,-1/2} \, ; \nonumber \\
\Delta L_{s,\,+1/2} & = & \Delta N_c - 2\sum_{\nu =1}^{\infty}
\nu\,\Delta N_{s,\,\nu} - \Delta L_{s,\,-1/2} \, .
\label{DScsNN}
\end{eqnarray}

For simplicity, let us assume again that the initial ground state
has zero-spin density $m=0$. In order to study the degree of
localization/delocalization of the quantum objects, we consider
elementary excitations associated with creation of $c$
pseudoparticles, $s,1$ pseudoparticles, $c,\nu$ pseudoparticles,
and $-1/2$ Yang holons. According to Eqs. (\ref{E1GS}) and
(\ref{om0}), for initial ground states with finite spin density
$m$ in the range $0<m<n$ the energy spectrum of excitations
involving creation of $s,\nu$ pseudoparticles belonging to $\nu>1$
branches and/or $-1/2$ Yang holons is gapped. The study of the
electron double occupation of such finite-spin-density excitations
is an interesting problem which will be considered elsewhere.

We start by considering the creation of a $c$ pseudoparticle at
band-momentum $q$. Creation of a $c$ pseudoparticle is associated
with creation of a spin-up electron, as confirmed by Eqs.
(\ref{DNu}) and (\ref{DNd}). From Eq. (\ref{DScsNN}) we find that
such an excitation also involves annihilation of a $+1/2$ Yang
holon and creation of a $+1/2$ HL spinon. According to Eqs.
(\ref{DDUi}) and (\ref{DlU}), in the limit $U/t\rightarrow\infty$
this elementary excitation leads to no change in the ground-state
electron double occupation $D_0$, Eq. (\ref{D0}). Therefore, in
this limit creation of a $c$ pseudoparticle corresponds to
creation of an extra electron singly occupied site and
annihilation of an empty side of the real-space lattice. For
decreasing values of $U/t$, the $U/t$ dependence of the $c$
pseudoparticle electron double occupation of Figs. 1 and 2 reveals
that this excitation leads to a small positive finite electron
double-occupation deviation $\Delta D=D_c (q)$. At both $q=0$ and
$q=\pm 2k_F$ this deviation is maximum when $U/t\rightarrow 0$,
where from Eq. (\ref{DU0}) and Figs. 1 and 2 we find that $\Delta
D=D_c(0)=D_c(\pm 2k_F) = n/2$. The deviation $\Delta D =D_c (q)$
changes slowly with band-momentum $q$ for all values of $U/t$ and
electronic density $n$. We emphasize that the kinetic-energy
deviation $\Delta T$ which results from this excitation is always
finite. We thus conclude that the $c$ pseudoparticle does not
describe localized electron configurations. The $c$ pseudoparticle
has it highest delocalization degree in the limit of
$U/t\rightarrow\infty$, where it leads to no electron
double-occupation deviation. In this case it becomes the
non-interacting spin-less fermion of Refs.
\cite{Ogata,Ricardo,Penc95,Penc96}.

Let us now consider the case of the $s,1$ pseudoparticle. Creation
of a $s,1$ pseudoparticle at band-momentum $q$ is associated with
a spin flip which creates a spin-down electron and annihilates a
spin-up electron. This is confirmed by analysis of Eqs.
(\ref{DNu}) and (\ref{DNd}). Moreover, it follows from Eq.
(\ref{DScsNN}) that creation of a $s,1$ pseudoparticle also
involves annihilation of two $+1/2$ HL spinons. The corresponding
electron double occupation deviation is given by $\Delta D =
D_{s,\,1}(q)$ and equals the $s,1$ double-occupation spectrum.
According to Eqs. (\ref{ineqD}), (\ref{DU0}), and (\ref{DlU}),
this spectrum has its minimum value at the band-momentum values
$q=\pm k_F$, where it vanishes. In addition, we find that at this
band-momentum value the $s,1$ pseudoparticle does not carry
kinetic energy. Thus at $q=\pm k_F$ and for all finite values of
$U/t$ and $n$ creation of the $s,1$ pseudoparticle is associated
with an on-site electronic spin flip process in the real-space
lattice. It follows that at this band-momentum values the $s,1$
pseudoparticle has a localized character. For decreasing values of
$\vert q\vert$ and finite values of $U/t$, the value of
$D_{s,\,1}(q)$ increases and we find that the $s,1$ pseudoparticle
carries kinetic energy. It follows that for $\vert q\vert < k_F$,
the spin flip electronic process has no pure on-site character.
The delocalization degree of the $s,1$ pseudoparticle is maximum
for band-momentum $q=0$. In contrast to the $c$ pseudoparticles
and $c,\nu$ pseudoparticles, at finite values of $U/t$ the degree
of delocalization of the $s,1$ pseudoparticle is in general
highest when the value of $D_{s,\,1}(q)$ is maximum. This can be
understood by considering the $U/t\rightarrow\infty$ limit. In
this case the $q=\pm k_F$ physics is extended to all values of $q$
and the $s,1$ pseudoparticle carries no kinetic energy. Moreover,
according to Eqs. (\ref{DDUi}) and (\ref{DlU}) the electron
double-occupation spectrum is in this limit given by
$D_{s,\,1}(q)= 0$. As was mentioned before, creation of the $s,1$
pseudoparticle at any value of $q$ for $U/t\rightarrow\infty$ is
associated with an on-site electronic spin-flip process. It
follows that in the case of the $s,1$ pseudoparticle the highest
degree of localization is achieved when it describes electron
singly-occupied site excitations only. Thus, as the value of the
on-site repulsion $U/t$ is decreased and $\Delta D =D_{s,\,1}(q)$
becomes finite, the degree of delocalization increases. This
effect is stronger for low values of band-momentum $q$. For the
limiting values $q=\pm k_F$ the $s,1$ pseudoparticle describes a
localized excitation for all values of $U/t$ and $n$, as mentioned
above. The curves plotted in Fig. 3 show that $\Delta D =
D_{s,\,1}(0)$ is maximum for an intermediate value of $U/t$.
Interestingly, in the limit of $U/t\rightarrow 0$ one finds that
$D_{s,\,1}(q)=0$ for all values of $q$, as in the limit
$U/t\rightarrow\infty$. However, in this case the $s,1$
pseudoparticle carries finite kinetic energy for all band-momentum
values $q$ except at $q=\pm k_F$. This means that for vanishing
values of $U/t$ the delocalization degree is not directly related
to electron double-occupation.

Another important elementary quantum object is the $-1/2$ Yang
holon. Creation of a $-1/2$ Yang holon is associated with creation
of two electrons of opposite spin projection, as is confirmed by
Eqs. (\ref{DNu}) and (\ref{DNd}). From Eq. (\ref{DScsNN}) we find
that creation of a $-1/2$ Yang holon implies annihilation of a
$+1/2$ Yang holon. The $-1/2$ Yang holon does not carry kinetic
energy and from Eq. (\ref{Dhs}) we find that its creation leads to
a electron double-occupation deviation value $\Delta D=1$.
Therefore, creation of a $-1/2$ Yang holon is associated for all
finite values of $U/t$ with the creation of a new localized
electron doubly occupied site and the annihilation of an empty
site in all real-space lattice electron site distribution
configurations of the initial ground state, as mentioned
previously. This is fully confirmed by the expression in terms of
electronic operators of the $\eta$-spin flip generators given in
Eq. (\ref{Sc}).

Let us now consider the $c,\nu$ pseudoparticle. This is a $s_c=0$
composite quantum object which results from the combination of
$\nu$ $-1/2$ holons and $\nu$ $+1/2$ holons \cite{I}. Since each
$-1/2$ holon is made out of two rotated electrons of opposite spin
projection \cite{I,III}, it follows that creation of a $c,\nu$
pseudoparticle at band-momentum $q$ is associated with creation of
a number $\nu$ of spin-down electrons and creation of an equal
number of spin-up electrons. (We note that the number of rotated
electrons equals that of electrons.) This is confirmed by Eqs.
(\ref{DNu}) and (\ref{DNd}). According to Eq. (\ref{DScsNN}),
creation of a $c,\nu$ pseudoparticle also implies annihilation of
$2\nu$ $+1/2$ Yang holons. The electron double-occupation
deviation originated by creation of a $c,\nu$ pseudoparticle is
given by $\Delta D = D_{c,\,\nu}(q)$. According to Eqs.
(\ref{ineqD}), (\ref{DU0}), and (\ref{DlU}) and Figs. 4 and 5,
this deviation has its maximum value for the band-momentum values
$q=\pm [\pi-2k_F]$, where it is given by $\nu$. In addition, at
this band-momentum values the $c,\nu$ pseudoparticle does not
carry kinetic energy. It follows that at these values of $q$
creation of the $c,\nu$ pseudoparticle is associated with creation
of $\nu$ localized electron double occupied sites and annihilation
of $\nu$ electron empty sites in all real-space lattice electron
site distribution configurations of the initial ground state. Note
that this result is valid for all electronic densities $n$ and
finite values of the on-site repulsion $U/t$. For these
band-momentum values the $c,\nu$ pseudoparticle corresponds to
localized real-space lattice electron site distribution
configurations similar to the ones obtained by creation of $\nu$
$-1/2$ Yang holons. For decreasing values of $\vert q\vert$ and
finite values of $U/t$ the value $D_{c,\,\nu}(q)$ decreases and we
find that the $c,\nu$ pseudoparticle carries finite kinetic
energy. Thus, in contrast to the case of the $s,1$ pseudoparticle,
in the present case the degree of delocalization increases with
the decreasing of the electron double-occupation deviation value.
For finite values of $U/t$ and band-momentum values $\vert q\vert
< [\pi-2k_F]$ the $2\nu$ created electrons are characterized by
some degree of delocalization which is maximum at $q=0$. This
means that for these band-momentum values the $2\nu$ electrons
have some degree of delocalization and are not all created in
pairs at the same site of the real-space lattice. On the other
hand, as $U/t\rightarrow\infty$ the $q=\pm [\pi-2k_F]$ physics is
extended to all values of $q$. In this limit the $c,\nu$
pseudoparticle does not carry kinetic energy for all values of $q$
and according to Eq. (\ref{DlU}) the electron double-occupation
spectrum is in that limit given by $D_{c,\,\nu}(q)=\nu$. As
discussed above, in this limit creation of a $c,\nu$
pseudoparticle for all values of $q$ is associated with creation
of $\nu$ localized electron double occupied sites and annihilation
of $\nu$ electron empty sites in all real-space lattice electron
site distribution configurations of the initial ground state.
Finally we emphasize that according to the total-momentum
expression (36) of Ref. \cite{I}, the $c,\nu$ pseudoparticle
momentum spectrum is $(\pi -q)$. Thus the band-momentum $q=0$ is
associated with a contribution to the momentum given by $(\pi
-q)=\pi$. This reveals that the maximum degree of delocalization
of the $c ,\nu$ pseudoparticles corresponds to a finite excitation
momentum of $\pi$. Note also that for all the remaining
pseudoparticle branches the momentum spectrum equals the
band-momentum $q$.

%%%%%%%%%%%%%%%%%%%%%%%%%%%%%%%%%%%%%%%%%%%%%%%%%%%%%%%%%%%%%%%%
\subsection{$U/t\rightarrow\infty$ ELECTRON DOUBLE-OCCUPATION SELECTION RULES}

Let the operator ${\hat{O}}_{\cal{N}}$ be a product of a finite
number

\begin{equation}
{\cal{N}}=\sum_{l_c,\,l_s=\pm 1} {\cal{N}}_{l_c,\,l_s} \, ,
\label{d}
\end{equation}
of one-electron creation and/or annihilation operators. Here
${\cal{N}}/N_a$ is vanishing small in the present thermodynamic
limit and ${\cal{N}}_{l_c,\,l_s}$ is the number of electronic
creation and annihilation operators for $l_c=-1$ and $l_c=+1$
respectively, and with spin down and spin up for $l_s=-1$ and
$l_s=+1$ respectively.

Let us consider states ${\hat{O}}_{{\cal{N}}}\vert GS\rangle$
which result from application of a general ${\cal{N}}$-electron
operator ${\hat{O}}_{{\cal{N}}}$ onto a ground state $\vert
GS\rangle$ or onto any eigenstate of the spin $\sigma$ electron
numbers. The pseudoparticle, holon, and spinon numbers of the
ground state are given in Eqs. (C24) and (C25) of Ref. \cite{I}.
We consider transformations generated by application of a
${\cal{N}}$-electron operator ${\hat{O}}_{{\cal{N}}}$ on any
eigenstate of the spin $\sigma$ electron numbers and not
necessarily on the ground state. Let $\vert\phi\rangle$ be such a
general state. It is not required that such an arbitrary state be
an energy eigenstate. The corresponding state
${\hat{O}}_{{\cal{N}}}\vert\phi\rangle$ is also an eigenstate of
the spin $\sigma=\uparrow ,\downarrow$ electron number operators.
It belongs to an electron ensemble space whose electron numbers
differ from the numbers of the initial state by deviations $\Delta
N_{\uparrow}$ and $\Delta N_{\downarrow}$, such that $\Delta N =
\Delta N_{\uparrow} +\Delta N_{\downarrow}$. These deviations can
be expressed in terms of the numbers ${\cal{N}}_{l_c,\,l_s}$ of
Eq. (\ref{d}) as follows,

\begin{equation}
\Delta N_{\uparrow} = \sum_{l_c=\pm 1}(-l_c)\,{\cal{N}}_{l_c,\,+1}
\, ; \hspace{1cm} \Delta N_{\downarrow} = \sum_{l_c=\pm
1}(-l_c)\,{\cal{N}}_{l_c,\,-1} \, .
\label{Nupdodll}
\end{equation}

Following the results of Ref. \cite{I}, let us consider the four
expectation values $R_{\alpha,\,l_{\alpha}}=\langle
{\hat{R}}_{\alpha,\,l_{\alpha}}\rangle$, where $\alpha =c,s$ and
$l_{\alpha}=-1,\,+1$ and the corresponding operator
${\hat{R}}_{c,\,-1}$ counts the number of electron doubly-occupied
sites, ${\hat{R}}_{c,\,+1}$ counts the number of electron empty
sites, ${\hat{R}}_{s,\,-1}$ counts the number of spin-down
electron singly-occupied sites, and ${\hat{R}}_{s,\,+1}$ counts
the number of spin-up electron singly-occupied sites. The
expressions of these operators are given in Eqs. (23) and (24) of
Ref. \cite{I}. These operators obey the following relations

\begin{equation}
{\hat{R}}_{c,\,+1} = N_a - {\hat{N}} + {\hat{R}}_{c,\,-1} \, ;
\hspace{1cm} {\hat{R}}_{s,\,-1} = {\hat{N}}_{\downarrow} -
{\hat{R}}_{c,\,-1} \, ; \hspace{1cm} {\hat{R}}_{s,\,+1} =
{\hat{N}}_{\uparrow} - {\hat{R}}_{c,\,-1} .
\label{RcsD}
\end{equation}
From the summation of the three relations of Eq. (\ref{RcsD}) we
find the following fourth dependent relation

\begin{equation}
\sum_{\alpha =c,s}\sum_{l_{\alpha}=\pm
1}{\hat{R}}_{\alpha,\,l_{\alpha}} = N_a \, .
\label{sumRNa}
\end{equation}

The operational relations (\ref{RcsD})-(\ref{sumRNa}) are valid
for the whole parameter space. Furthermore, they reveal that at
given spin $\sigma$ electron numbers, out of the four expectation
values of the operators of Eq. (\ref{sumRNa}) only one is
independent. In the previous section we have studied the
double-occupation expectation value $D\equiv R_{c,\,-1}$. The use
of the operational Eqs. (\ref{RcsD})-(\ref{sumRNa}) provides the
corresponding values for $R_{c,\,+1}$, $R_{s,\,-1}$, and
$R_{s,\,+1}$.

In the limit $U/t\rightarrow\infty$ the energy-eigenstate
expectation values of the four number operators
${\hat{R}}_{c,\,-1}$, ${\hat{R}}_{c,\,+1}$, ${\hat{R}}_{s,\,-1}$,
and ${\hat{R}}_{s,\,+1}$ become good quantum numbers \cite{I,III}.
Thus in that limit each energy eigenstate corresponds to
real-space lattice electron site distribution configurations with
the same values for $R_{c,\,-1}$, $R_{c,\,+1}$, $R_{s,\,-1}$, and
$R_{s,\,+1}$. For finite values of $U/t$ each energy eigenstate
corresponds in general to a superposition of real-space lattice
electron site distribution configurations with different numbers
of electron doubly-occupied sites, electron empty sites, spin-down
electron singly-occupied sites, and spin-up electron
singly-occupied sites.

Let us use the relations of Eq. (\ref{RcsD}) to evaluate
expressions for the expectation values $R_{c,\,+1}$, $R_{s,\,-1}$,
and $R_{s,\,+1}$ for the particular case of a ground state, from
the corresponding $m=0$ electron double-occupation expression
defined by Eqs. (\ref{D0}) and (\ref{flim}). We find that for
$m=0$ and for all values of $U/t$ and for values of the electronic
density in the domain $0\leq n\leq 1$, these ground-state
expectation values read,

\begin{equation}
R^0_{c,\,+1} = N_a - N + {N\over 2}\,\Bigl({n\over
2}\Bigr)\,f(n,U/t) \, ,
\label{R0c1}
\end{equation}
and

\begin{equation}
R^0_{s,\,-1} = R^0_{s,\,+1} = {N\over 2}\Bigl[1 - \Bigl({n\over
2}\Bigr)\,f(n,U/t)\Bigr] \, ,
\label{R0s-1}
\end{equation}
respectively. Here $f(n,U/t)$ is the function given in Eqs.
(\ref{flim}) and (\ref{fn1}). Note that these ground-state
expectation values change from $R^0_{c,\,-1}\equiv D_0={N\over
2}\,\Bigl({n\over 2}\Bigr)$, $R^0_{c,\,+1}= [N_a - N] + {N\over
2}\,\Bigl({n\over 2}\Bigr)$, and
$R^0_{s,\,-1}=R^0_{s,\,+1}={N\over 2} \Bigl[1 - \Bigl({n\over
2}\Bigr)\Bigr]$ as $U/t\rightarrow 0$ to $R^0_{c,\,-1}\equiv
D_0=0$, $R^0_{c,\,+1}= [N_a - N]$, and $R^0_{s,\,-1}= R^0_{s,\,+1}
={N\over 2}$ when $U/t\rightarrow\infty$ and these quantities
become good quantum numbers.

Importantly, in the limit $U/t\rightarrow\infty$ the four
deviations $\Delta
R_{\alpha,\,l_{\alpha}}=[R_{\alpha,\,l_{\alpha}}-R^0_{\alpha,\,l_{\alpha}}]$
such that $\alpha =c,s$ and $l_{\alpha}=-1,+1$ of any eigenstate
of the spin $\sigma=\uparrow ,\downarrow$ electron number
operators which result from application onto that state of the
above general operator ${\hat{O}}_{\cal{N}}$ are restricted to the
following ranges,

\begin{equation}
-\sum_{l_s=\pm 1} {\cal{N}}_{+1,\,l_s}\leq\Delta D\equiv \Delta
R_{c,\,-1}\leq\sum_{l_s=\pm 1} {\cal{N}}_{-1,\,l_s} \, ,
\label{DRc-range}
\end{equation}

\begin{equation}
-\sum_{l_s=\pm 1} {\cal{N}}_{-1,\,l_s}\leq\Delta
R_{c,\,+1}\leq\sum_{l_s=\pm 1} {\cal{N}}_{+1,\,l_s} \, ,
\label{DRc+range}
\end{equation}

\begin{equation}
-\sum_{l_c,l_s=\pm 1}\delta_{l_c,\,-l_s}\,
{\cal{N}}_{l_c,\,l_s}\leq\Delta R_{s,\,-1}\leq\sum_{l_c,l_s=\pm
1}\delta_{l_c,\,l_s}\, {\cal{N}}_{l_c,\,l_s} \, ,
\label{DRs-range}
\end{equation}
and

\begin{equation}
-\sum_{l_c,l_s=\pm 1}\delta_{l_c,\,l_s}\,
{\cal{N}}_{l_c,\,l_s}\leq\Delta R_{s,\,+1}\leq\sum_{l_c,l_s=\pm
1}\delta_{l_c,\,-l_s}\, {\cal{N}}_{l_c,\,l_s} \, ,
\label{DRs+range}
\end{equation}
respectively. Obviously, given one of the four ranges of values
defined by the inequalities (\ref{DRc-range})-(\ref{DRs+range}),
the other three follow from the operational relations of Eq.
(\ref{RcsD}).

The maximum (and minimum) value of the inequalities
(\ref{DRc-range}) and (\ref{DRc+range}) (and (\ref{DRs-range}) and
(\ref{DRs+range})) are reached when all the electrons created by
the operator ${\hat{O}}_{\cal{N}}$, in a number of $\sum_{l_s=\pm
1} {\cal{N}}_{-1,\,l_s}$, transform an equal number of electron
singly-occupied sites into electron doubly-occupied sites. In
addition, it is required that all the annihilation operators of
that operator, in a number of $\sum_{l_s=\pm 1}
{\cal{N}}_{+1,\,l_s}$, transform an equal number of electron
singly-occupied sites into electron empty sites for all the
real-space lattice electron site distribution configurations of
the initial state with a number of electron singly occupied sites
larger than or equal to ${\cal{N}}$. Note that if the initial
state includes distribution configurations whose number of
electron singly occupied sites is smaller than ${\cal{N}}$, then
the maximum value of $\Delta D$ is smaller than $\sum_{l_s=\pm 1}
{\cal{N}}_{-1,\,l_s}$ and thus the inequalities (\ref{DRc-range})
and (\ref{DRc+range}) (and (\ref{DRs-range}) and
(\ref{DRs+range})) are also satisfied.

The minimum (and maximum) value of the inequalities
(\ref{DRc-range}) and (\ref{DRc+range}) (and (\ref{DRs-range}) and
(\ref{DRs+range})) are reached when all the electrons created by
the operator ${\hat{O}}_{\cal{N}}$, in number of $\sum_{l_s=\pm 1}
{\cal{N}}_{-1,\,l_s}$, transform an equal number of electron empty
sites into electron singly-occupied sites. Furthermore, it is
required that all the $\sum_{l_s=\pm 1} {\cal{N}}_{+1,\,l_s}$
annihilation operators of that operator transform an equal number
of electron doubly-occupied sites into electron singly-occupied
sites for all the real-space lattice electron site distribution
configurations of the initial state with a number of electron
empty sites equal to or larger than $\sum_{l_s=\pm 1}
{\cal{N}}_{-1,\,l_s}$ and a number of electron doubly-occupied
sites equal to or larger than $\sum_{l_s=\pm 1}
{\cal{N}}_{+1,\,l_s}$. Note that if the initial state includes
distribution configurations whose number of electron empty sites
is smaller than $\sum_{l_s=\pm 1} {\cal{N}}_{-1,\,l_s}$ and/or
whose number of electron doubly-occupied sites is smaller than
$\sum_{l_s=\pm 1} {\cal{N}}_{+1,\,l_s}$, the minimum value of
$\Delta D$ is larger than $-\sum_{l_s=\pm 1} {\cal{N}}_{+1,\,l_s}$
and thus the inequalities (\ref{DRc-range}) and (\ref{DRc+range})
[and (\ref{DRs-range}) and (\ref{DRs+range})] are also satisfied.

It is found in Ref. \cite{III} that the $U/t\rightarrow\infty$
electron double-occupation selection rules associated with the
inequalities (\ref{DRc-range})-(\ref{DRs+range}) are related to
the occurrence of other selection rules which concern the range of
the deviations in the number of $\pm 1/2$ holons and $\pm 1/2$
spinons. The holon and spinon selection rules are valid for all
values of $U/t$ and play an important role in the evaluation of
correlation functions \cite{III,IV,V}. These selection rules are
exact in the case of rotated-electron operators. In the case of
the corresponding electron operators such rules provide the
numbers of holons and spinons of the final states which contribute
most significantly to the few-electron correlation functions
\cite{III}. This relationship between the $U/t\rightarrow\infty$
electron double-occupation deviations and the deviations in the
number $\pm 1/2$ holons and $\pm 1/2$ spinons follows from the
unitary transformation which maps electrons onto rotated
electrons. This transformation also maps the electron
double-occupation operator onto an operator that counts the number
of $-1/2$ holons. The latter operator is nothing but the
rotated-electron double occupation operator \cite{III}.

%%%%%%%%%%%%%%%%%%%%%%%%%%%%%%%%%%%%%%%%%%%%%%%%%%%%%%%%%%%%%%%%%%%%%%%%%%
\section{THE PSEUDOPARTICLE ENERGY BANDS}

In section IV we found that the pseudoparticle energy bands
defined by Eqs. (C15)-(C21) of Ref. \cite{I} play an important
role in the electron double occupation spectra of the
pseudoparticles, as confirmed by Eq. (\ref{barepsi}). Moreover, in
reference \cite{V} it is found that the shape of the lines in the
frequency/energy and momentum plane where the peaks or edges in
the one-electron and two-electron spectral-weight distributions
are located corresponds to the pseudoparticle energy bands. Thus
the study of the band-momentum dependence of these energy bands is
a problem of interest for the understanding of the one-electron
and two-electron spectral properties of the quantum liquid. The
$c$ pseudoparticle and $s,1$ pseudoparticle energy bands were
studied and plotted in Ref. \cite{Carmelo91} for finite values of
the spin density. On the other hand, the $c,\nu$ pseudoparticle
energy bands and $s,\nu$ pseudoparticle energy bands for $\nu>1$
were first introduced in Ref. \cite{Carmelo97}, yet these bands
were not plotted in that reference. Since most studies on the
finite-energy spectral properties of low-dimensional materials
refer to zero magnetization, it is worthwhile plotting and
discussing the band-momentum dependence of the above
pseudoparticle bands for zero spin density.

The general excitation energy spectrum of interest for the problem
of the finite-energy correlation functions is given in Eq.
(\ref{E1GS}). The band-momentum independent term $\omega_0$ is
defined in Eq. (\ref{om0}). The value of this finite energy is
determined by the holon and spinon deviations and correspond to
the finite-energy edges of the correlation functions. The other
term of the energy spectrum (\ref{E1GS}) is of gapless character
and in the case of zero magnetization involves the pseudoparticle
energy bands $\epsilon_{c}(q)$, $\epsilon_{s,\,1}(q)$, and
$\epsilon_{c,\,\nu}^0(q)$ where $\nu=1,2,3,...$. These bands are
defined in Eqs. (C15)-(C17) of Ref. \cite{I}. As discussed below,
at zero magnetization both the energy and band-momentum width of
the energy bands $\epsilon_{s,\,\nu}^0(q)$ vanishes in the case of
the $\nu >1$ branches. According to Eqs. (C1)-(C3) of Ref.
\cite{I}, in the case of the zero-spin-density ground states the
band $\epsilon_{c}(q)$ is filled (and empty) for band-momentum
values such that $\vert q\vert\leq 2k_F$ (and $2k_F<\vert
q\vert\leq\pi$). On the other hand, for such ground states the
band $\epsilon_{s,\,1}(q)$ is full and the bands
$\epsilon_{c,\,\nu}^0(q)$ where $\nu=1,2,3,...$ are empty. The
expressions of the bands $\epsilon_{c}(q)$, $\epsilon_{s,\,1}(q)$,
$\epsilon_{c,\,1}^0(q)$, and $\epsilon_{c,\,2}^0(q)$ are plotted
in Figs. 6-9 as a function of the band-momentum $q$ for
spin-density $m=0$ and different values of $U/t$ and of the
electronic density $n$.

As is confirmed by analysis of the energy-band curves plotted in
Figs. 6 (a) and 6 (b), both the energy and band-momentum width of
the $c$ pseudoparticle bands are density, spin-density, and $U/t$
independent and given by $4t$ and $2\pi$ respectively. From the
$U/t$ dependence of the curves of these figures, we find that the
main effect of increasing the on-site repulsion $U/t$ is the
increasing of the energy-band width of the ground-state filled sea
and the corresponding decreasing of the band-energy width of the
corresponding unfilled region. The main effect of changing the
electronic density is on the ground-state $c$ pseudoparticle {\it
Fermi} points, which are given by $\pm 2k_F$.

In contrast to the energy and band-momentum widths of the $c$
pseudoparticles, the energy width of the $\alpha ,\nu$
pseudoparticle bands plotted in Figs 7-9 is a decreasing function
of $U/t$, vanishing in the limit $U/t\rightarrow\infty$. This
property is a result of the fact that all $\eta$-spin and spin
configurations are degenerated in that limit. The band-momentum
width of the $c,\nu$ pseudoparticle bands and of the $s,\nu$
pseudoparticle bands belonging to branches such that $\nu>1$ are
density and spin-density dependent respectively, and given by
$2[\pi -2k_F]$ and $2[k_{F\uparrow}-k_{F\downarrow}]$
respectively. Note also that the $\alpha ,\nu$ pseudoparticle
energy bands $\epsilon_{s,\,1}^0(q)=\epsilon_{s,\,1}(q)-2\mu_0\,H$
and $\epsilon_{\alpha,\,\nu}^0(q)$ vanish at the band-momentum
limiting values, {\it i.e.}

\begin{equation}
\epsilon_{\alpha,\,\nu}^0(\pm q_{\alpha,\,\nu}) = 0 \, ,
\label{eql=0}
\end{equation}
where according to Eq. (C12)-(C14) of Ref. \cite{I},
$q_{c,\,\nu}=[\pi -2k_F]$, $q_{s,\,1}=k_{F\uparrow}$, and
$q_{s,\,\nu}=[k_{F\uparrow}-k_{F\downarrow}]$ for the $\nu >1$
branches.

For finite values of the spin density $m$ and magnetic field $H$,
the $s,\nu$ pseudoparticle energy bands $\epsilon^0_{s,\,\nu} (q)$
have for the $\nu >1$ branches both finite momentum width, given
by the above expression $2[k_{F\uparrow}-k_{F\uparrow}]$, and
finite energy with, given by $\vert \epsilon^0_{s,\,\nu}
(0)\vert$. The energy band value at $q=0$ is negative, {\it i.e.}
$\epsilon^0_{s,\,\nu} (0)\leq 0$. As the spin density
$m\rightarrow 0$, both widths
$2[k_{F\uparrow}-k_{F\uparrow}]\rightarrow 0$ and $\vert
\epsilon^0_{s,\,\nu} (0)\vert\rightarrow 0$ vanish. Thus, for the
$\nu>1$ branches the $s,\nu$ pseudoparticle energy bands collapse
into the point $\{q=0\,,\epsilon_{s,\,\nu}^0 (0)=0\}$ in that
limit. Since Figs. 6-9 correspond to $m=0$, these bands are not
plotted in these figures. The finite-magnetization case will be
considered elsewhere.

The $c,\nu$ pseudoparticle bands plotted in Figs. 8 (a)-(b) and 9
(a)-(b) have their maximum value at $q=0$. These bands are thus
inverted. As discussed below, as the electronic density approaches
$1$, both the energy and momentum widths of these bands vanish.
Except for the case of the $s,1$ band of Fig. 7, all remaining
bands plotted in Figs. 6-9 show for finite values of $U/t$ a
quadratic dependence on $q$ in the vicinity of the band-momentum
limiting values $q=\pm q_c=\pm\pi$ and $q=\pm q_{c,\,\nu}=\pm
[\pi-2k_F]$. In contrast, note that the $s,1$ band has a linear
$q$ dependence in the vicinity of the band-momentum limiting
values $q=\pm q_{s,\,1}=\pm k_F$. On the other hand, for finite
values of $U/t$ all pseudoparticle bands show a quadratic
dependence on the band-momentum $q$ in the vicinity of the point
$q=0$, as confirmed by the curves of Figs. 6-9.

From analysis of Figs. 8 and 9, we find that the
$\epsilon_{c,\,1}^0(q)$ and $\epsilon_{c,\,2}^0(q)$ bands vanish
at the band-momentum limiting values $q=\pm q_{c,\,\nu}=\pm [\pi
-2k_F]$. This behavior also occurs for finite values of the spin
density $m$. For such a values also the $\epsilon_{s,\,\nu}^0(q)$
bands vanish for $q=\pm q_{s,\,\nu}=\pm
[k_{F\uparrow}-k_{F\downarrow}]$. The energy band
$\epsilon^0_{c,\,\nu} (q)$ has its maximum value at $q=0$, with
$\epsilon^0_{c,\,\nu} (0)\geq 0$, as shown in Figs. 8 (a)-(b) and
9 (a)-(b). The energy band $\epsilon^0_{s,\,\nu} (q)$, for spin
density $m>0$ has its minimum at $q=0$, with $\epsilon^0_{s,\,\nu}
(0)\leq 0$, as in the case of the $s,1$ energy band plotted in
Fig. 7 for $m=0$.

In the Appendix we consider the limiting expressions of the
pseudoparticle energy bands in the case of zero spin density,
$m=0$. Note that according to the expressions given in Eqs.
(\ref{limec})-(\ref{limecn}) of the Appendix, the $q$ dependence
of the $c$ pseudoparticle energy band in the vicinity of
$q=\pm\pi$ and of the $c,\nu$ bands in the vicinity of $q=0$ is
different for finite values of $U/t$ and in the limit
$U/t\rightarrow 0$. As mentioned above, for finite values of $U/t$
the dependence on the band-momentum $q$ is quadratic in the
vicinity of these band-momentum values. Thus in that case the $q$
derivative of these bands vanishes both at $q=\pm\pi$ and $q=0$
respectively, as confirmed by the curves of Figs. 6-9 and by the
expressions of Eqs. (\ref{limec}) and (\ref{limecn}) of the
Appendix. In contrast, analysis of the data of these figures and
equations reveals that in the limit $U/t\rightarrow 0$ the $q$
derivative at these band-momentum values becomes finite and the
$q$ dependence in their vicinity is linear instead of quadratic.
This singular change results from the non-perturbative character
of the electronic correlations which leads to a different physics
at $U/t=0$ and for $U/t>0$. In the particular case of
half-filling, this effect occurs for the $c$ pseudoparticle band
at the band-momentum {\it Fermi points} $q=\pm\pi$. In this case,
that effect is associated with the singular Mott-Hubbard insulator
- metal transition. For other electronic densities this effect
occurs at unoccupied ground-state band-momentum values and
therefore, is not associated with such a transition. The vanishing
at half-filling of the band $\epsilon_{c,\,\nu}^0 (q)$ of Eq.
(\ref{ecnn1}) of the Appendix results from its collapse into the
point $\{q=0\,,\epsilon_{c,\,\nu}^0 (0)=0\}$ as $\delta\rightarrow
0$, where $\delta =(1-n)$ is the doping concentration. This is
confirmed by the dependence on $n$ (and $\delta$) of expressions
(\ref{limecn}) for that band. We emphasize that both the $s,\,\nu$
bands for the $\nu
>1$ branches and the $c,\nu$ bands collapse into the point
$\{q=0\,,\epsilon_{\alpha,\,\nu}^0 (0)=0\}$ with $\alpha =s$ and
$\alpha =c$ respectively, as the spin density $m\rightarrow 0$ and
the doping $\delta =(1-n)\rightarrow 0$ respectively.

The terms of the Landau energy functional given in Eq.
(\ref{E1GS}) describe the energy spectra of all one-electron and
two-electron elementary excitations of the 1D Hubbard model
\cite{I}. Within such a functional description, each elementary
excitation is simply described by specific values of the
pseudoparticle band-momentum distribution function deviations
defined in Eq. (58) of Ref. \cite{I}. These deviations describe:
(a) pseudoparticle - pseudoparticle hole processes in the $c$ and
$s,1$ bands which conserve the pseudoparticle numbers; (b)
creation or annihilation of $c$ and $s,1$ pseudoparticles; (c)
creation of $c,\nu$ pseudoparticles and of $s,\nu$ pseudoparticles
with $\nu>1$; and (d) creation of $-1/2$ Yang holons and of $-1/2$
LH spinons.

All the energy spectra obtained previously in the literature, as
for instance the ones studied in the Refs.
\cite{Ovchinnikov,Coll,Woynarovich,Choy,Klumper,Deguchi}, can be
expressed in terms of the pseudoparticle energy bands plotted for
$m=0$ in Figs. 6-9. The authors of these studies used for each
specific elementary excitation different forms for the two Lieb
and Wu equations \cite{Lieb} or/and for the thermodynamic
Takahashi's equations \cite{Takahashi} provided in Ref. \cite{I}.
On the other hand, insertion in the energy Landau-liquid
functional (\ref{E1GS}) of band-momentum distribution function
deviations suitable to these excitations leads to the same energy
spectra. In addition, the energy functional (\ref{E1GS}) provides
the spectrum for all values of the energy of all excitations which
contribute to the one-electron and two-electron physics, as
mentioned above.

%%%%%%%%%%%%%%%%%%%%%%%%%%%%%%%%%%%%%%%%%%%%%%%%%%%%%%%%%%%%%%%%
\section{CONCLUDING REMARKS}

In this paper we studied the electron double-occupation deviations
which result from creation of $c$ pseudoparticles, $\alpha,\nu$
pseudoparticles, $-1/2$ Yang holons, and $-1/2$ HL spinons, where
$\alpha =c,\,s$ and $\nu=1,2,3,...$. All energy eigenstates can be
described in terms of occupancy configurations of these quantum
objects which are related to electrons and rotated electrons in
Refs. \cite{I,III}. Here we introduced an electron
double-occupation functional whose coefficients are pseudoparticle
band-momentum dependent double-occupation spectra. Our
investigations provided interesting information about the electron
site distribution configurations of the real-space lattice which
describe the energy eigenstates. The study of these spectra
provided important information on the localization/deslocalization
degree of the real-space lattice electron site distribution
configurations which describe the elementary quantum objects. We
found that for some of the pseudoparticle branches such a degree
of localization/delocalization is strongly dependent on the value
of the band momentum. We also considered $U/t\rightarrow\infty$
selection rules, which limit the ranges of the electron
double-occupation deviations that result from excitations
generated by $\cal{N}$-electron physical operators. These rules
are the starting point for finding related selection rules, which
are investigated elsewhere \cite{III}. The latter rules are valid
for all values of $U/t$ and play an important role in the study of
finite-energy correlation functions.

This paper is a first step in the use of the general holon and
spinon scenario in the study of the finite-energy spectral
properties of the 1D Hubbard model. In addition to the
introduction and study of the electron double-occupation
functional, in this paper we studied the dependence on the
band-momentum $q$, on-site repulsion $U/t$, and electronic density
$n$ of the pseudoparticle energy bands associated with such a
functional. As was mentioned in this paper, these bands play a
central role in the spectral properties of the model \cite{V}.

The holon and spinon description of the 1D Hubbard model
introduced in Ref. \cite{I} and further studied here is used in
Ref. \cite{III} in the finding of further useful microscopic
information on the spectral properties of the model. By combining
the results obtained here with that information, one can evaluate
expressions for finite-energy one-electron and two-electron
correlation functions \cite{I,IV,V}. That is a problem of major
importance for the study of the unusual finite-energy spectral
properties observed in real quasi-one-dimensional materials
\cite{Hussey,Menzel,Fuji,Hasan,Ralph,Gweon,Monceau,Takenaka,Mizokawa,Moser,Mihaly,Vescoli00,Denlinger,Fujisawa,Kobayashi,Bourbonnais,Vescoli,Zwick,Mori,Kim}.
Elsewhere it is confirmed that the holon and spinon description of
the energy eigenstates introduced in Ref. \cite{I} and further
developed here and in the related articles \cite{I,III,IV,V} is
suitable for the successful study of these finite-energy spectral
properties.

%%%%%%%%%%%%%%%%%%%%%%%%%%%%%%%%%%%%%%%%%%%%%%%%%%%%%%%%%%%%%%%%%%%%%%%%%%
\begin{acknowledgments}
We thank Jim W. Allen, Daniel Bozi, Ant\^onio Castro Neto, Ralph
Claessen, Francisco (Paco) Guinea, Katrina Hibberd, Peter Horsch,
Jo\~ao M. B. Lopes dos Santos, Lu\'{\i}s M. Martelo, Karlo Penc,
and J. M. Rom\'an for stimulating discussions.
\end{acknowledgments}
%%%%%%%%%%%%%%%%%%%%%%%%%%%%%%%%%%%%%%%%%%%%%%%%%%%%%%%%%%%%%%%%%%%%%%%%%%
\appendix

%%%%%%%%%%%%%%%%%%%%%%%%%%%%%%%%%%%%%%%%%%%%%%%%%%%%%%%%%%%%%%%%%%%
\section{LIMITING EXPRESSIONS OF THE PSEUDOPARTICLE ENERGY BANDS}

In this Appendix we present the limiting expressions of the
pseudoparticle energy bands in the case of zero spin density,
$m=0$. In the limits $U/t\rightarrow 0$ and $U/t >> 1$ one can
evaluate closed form expressions for the bands $\epsilon_c (q)$,
$\epsilon_{s,\,1} (q)$, and $\epsilon^0_{\alpha ,\,\nu} (q)$ where
$\alpha =c,\,s$ and $\nu=1,2,3,...$, defined in Appendix C of Ref.
\cite{I}. The simplest case is that of the $s,\nu$ energy bands
for the $\nu>1$ branches, which at zero spin density $m=0$ are
given for all values of $U/t$ and $n$ by $\epsilon_{s,\,\nu}^0 (q)
= 0$ for $\vert q\vert \leq (k_{F\uparrow} -k_{F\downarrow}) = 0$.
As $m\rightarrow 0$ these $\nu >1$ bands collapse in that limit
into the point $\{q=0\,,\epsilon_{s,\,\nu}^0 (0)=0\}$. The
expressions of the remaining pseudoparticle bands for zero spin
density $m=0$, values of the density $0\leq n\leq 1 $, and
$U/t\rightarrow 0$ and $U/t>>1$ are the following,

\begin{eqnarray}
\epsilon_c (q) & = & -4t\,\Bigl[\cos\Bigl({q\over 2}\Bigr)-\cos
\Bigl({\pi n\over 2}\Bigr)\Bigr] \, ; \hspace{0.5cm} \vert q\vert
\leq 2k_F \, , \hspace{0.5cm} U/t\rightarrow 0 \, ;  \nonumber \\
& = & -2t\,\Bigl[\cos\Bigl(\vert q\vert - {\pi n\over
2}\Bigr)-\cos \Bigl({\pi n\over 2}\Bigr)\Bigr] \, ; \hspace{1cm}
2k_F \leq \vert q\vert \leq \pi \, , \hspace{0.5cm} U/t\rightarrow
0 \, ; \nonumber \\ & = & -2t\,\Bigl[\cos (q)-\cos
(\pi n)\Bigr] \nonumber \\
& - & {(2t)^2\over U}\,2n\,\ln (2)\,\Bigl[\,\sin^2 (q)-\sin^2 (\pi
n) \Bigr] \, ; \hspace{0.5cm} \vert q\vert \leq \pi \, ,
\hspace{0.2cm} U/t
>> 1 \, ,
\label{limec}
\end{eqnarray}

\begin{eqnarray}
\epsilon_{s,\,1} (q) & = & -2t\,\Bigl[\cos (q) -\cos \Bigl({\pi
n\over 2}\Bigr)\Bigr] \, ; \hspace{0.5cm} \vert q\vert \leq k_F \,
, \hspace{0.5cm} U/t\rightarrow 0  \nonumber \\ & = &
-{(2t)^2\over U}\,{\pi\over 2}\,\Bigl[n - {\sin (2\pi n)\over
2\pi}\Bigr]\,\cos \Bigl({q\over n}\Bigr) \, ; \hspace{0.5cm} \vert
q\vert \leq k_F \, , \hspace{0.2cm} U/t >> 1 \, , \label{limes1}
\end{eqnarray}

\begin{eqnarray}
\epsilon^0_{c,\,\nu} (q) & = & 4t\,\cos\Bigl({\vert q\vert + \pi n
\over 2}\Bigr) \, ; \hspace{0.5cm}  \vert q\vert \leq (\pi-2k_F)
\, , \hspace{0.5cm} U/t\rightarrow 0  \nonumber \\ & = &
{(4t)^2\over U}\,{1\over 2\nu}\,\Bigl[\delta - {\sin
(2\pi\delta)\over 2\pi}\Bigr]\,\cos^2 \Bigl({q\over 2\delta}\Bigr)
\, ; \hspace{0.5cm} \vert q\vert \leq (\pi-2k_F) \, ,
\hspace{0.2cm} U/t >> 1 \, , \label{limecn}
\end{eqnarray}
where $\delta\equiv (1-n)$ is the doping concentration. The
expressions (\ref{limec}) and (\ref{limes1}) were already obtained
in the $c$ and $s\equiv s,1$ pseudoparticle studies of Ref.
\cite{Carmelo91}. From analysis of the band-momentum $q$, on-site
repulsion $U/t$, and density $n$ dependence of the band
expressions (\ref{limec})-(\ref{limecn}), we find that these agree
with the corresponding dependence on the same parameters of the
curves of Figs. 6-9.

At half filling and spin density $m=0$ the integral equations
which define the ground-state rapidities and pseudoparticle bands
can be solved by Fourier transform. Thus in this case one can
derive the following expressions for the pseudoparticle bands,
which are valid for all values of $U/t>0$ \cite{Carmelo92},

\begin{equation}
\epsilon_c (q) = -2t\Bigl[\cos k^0 (q)+1\Bigr]
-4t\,\int_{0}^{\infty}dx\,J_1(x)\,{\cos \Bigl(x\,\sin k^0
(q)\Bigr)\over x\,[1+e^{xU \over 2t}]} \, ; \hspace{0.5cm} \vert
q\vert \leq \pi  \, ,
\label{ecn1}
\end{equation}

\begin{equation}
\epsilon_{s,\,1} (q) = -2t\int_{0}^{\infty}dx\, J_1(x)\,{\cos
\Bigl(x\,\Lambda^0_{s,\,1} (q)\Bigr) \over x\,\cosh({xU\over 4t})}
\, ; \hspace{0.5cm} \vert q\vert \leq \pi/2 \, ,
\label{es1n1}
\end{equation}
and

\begin{equation}
\epsilon_{c,\,\nu}^0 (q) = 0 \, ; \hspace{1cm} \vert q\vert \leq
(\pi -2k_F) = 0 \, ,
\label{ecnn1}
\end{equation}
The inverse functions of the half-filling ground-state
rapidity-momentum function $k^0 (q)$ and rapidity function
$\Lambda^0_{s,\,1} (q)$ involved in expressions (\ref{ecn1}) and
(\ref{es1n1}) are given in Eqs. (\ref{k0n1}) and (\ref{Ls1n1})
respectively.

%%%%%%%%%%%%%%%%%%%%%%%%%%%%%%%%%%%%%%%%%%%%%%%%%%%%%%%%%%%%%%%%%%%%%%%%%%

%**********************************************************
%************** F I G U R E   C A P T I O N S *************
%**********************************************************

\section*{FIGURE CAPTIONS}

\label{fig1}Figure 1 - The electron double-occupation parameter
$D_c (0)$ plotted as a function of $U/t$ for different values of
the electronic density $n$.

\vspace{0.3cm}

\label{fig2}Figure 2 - The electron double-occupation parameter
$D_c (2k_F)$ plotted as a function of $U/t$ for different values
of the electronic density $n$.

\vspace{0.3cm}

\label{fig3}Figure 3 - The electron double-occupation parameter
$D_{s,\,1} (0)$ plotted as a function of $U/t$ for different
values of the electronic density $n$.

\vspace{0.3cm}

\label{fig4}Figure 4 - The electron double-occupation parameter
$D_{c,\,1} (0)$ plotted as a function of $U/t$ for different
values of the electronic density $n$.

\vspace{0.3cm}

\label{fig5}Figure 5 - The electron double-occupation parameter
$D_{c,\,2} (0)$ plotted as a function of $U/t$ for different
values of the electronic density $n$.

\vspace{0.3cm}

\label{fig6}Figure 6 - The pseudoparticle energy band $\epsilon_c
(q)$ in units of $t$ plotted for electronic density (a) $n=1/2$
and (b) $n=5/6$ and on-site repulsion $U/t\rightarrow 0$,
$U/t=2.0$, $U/t=4.0$, $U/t=10$, $U/t=20$, and
$U/t\rightarrow\infty$. The ground-state {\it Fermi} level
corresponds to zero energy and is marked by a horizontal line
which overlaps the $\epsilon_c (q)$ band at the band-momentum {\it
Fermi} points $q=\pm 2k_F=\pm \pi/2$.

\vspace{0.3cm}

\label{fig7}Figure 7 - The pseudoparticle energy band
$\epsilon_{s,\,1} (q)$ in units of $t$ plotted for electronic
densities $n=1/2$ and $n=5/6$ and on-site repulsion
$U/t\rightarrow 0$, $U/t=2.0$, $U/t=4.0$, $U/t=10$, $U/t=20$, and
$U/t\rightarrow\infty$. The ground-state {\it Fermi} level
corresponds to zero energy overlaps the energy band at the
band-momentum {\it Fermi} points $q=\pm k_F$.

\vspace{0.3cm}

\label{fig8}Figure 8 - The pseudoparticle energy bands (a)
$\epsilon^0_{c,\,1} (q)$ and (b) $\epsilon^0_{c,\,2} (q)$ in units
of $t$ plotted for electronic density $n=1/2$ and on-site
repulsion $U/t\rightarrow 0$, $U/t=2.0$, $U/t=4.0$, $U/t=10$, and
$U/t=20$. The $U/t\rightarrow\infty$ limit of these bands
corresponds to a horizontal line located at the zero energy level.

\vspace{0.3cm}

\label{fig9}Figure 9 - The pseudoparticle energy bands (a)
$\epsilon^0_{c,\,1} (q)$ and (b) $\epsilon^0_{c,\,2} (q)$ in units
of $t$ plotted for electronic density $n=5/6$ and on-site
repulsion $U/t\rightarrow 0$, $U/t=2.0$, $U/t=4.0$, $U/t=10$, and
$U/t=20$. The $U/t\rightarrow\infty$ limit of these bands
corresponds to a horizontal line located at the zero energy level.
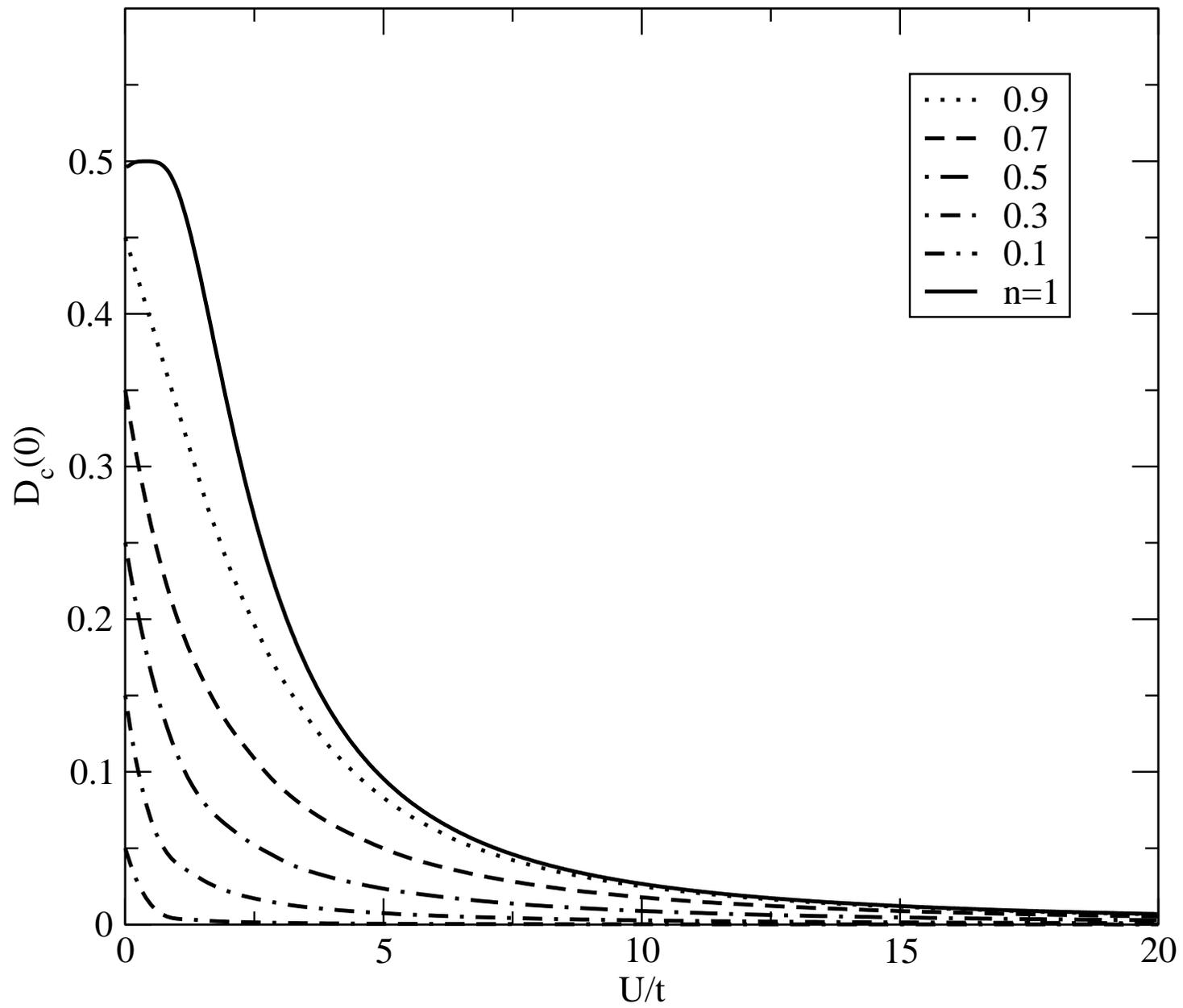
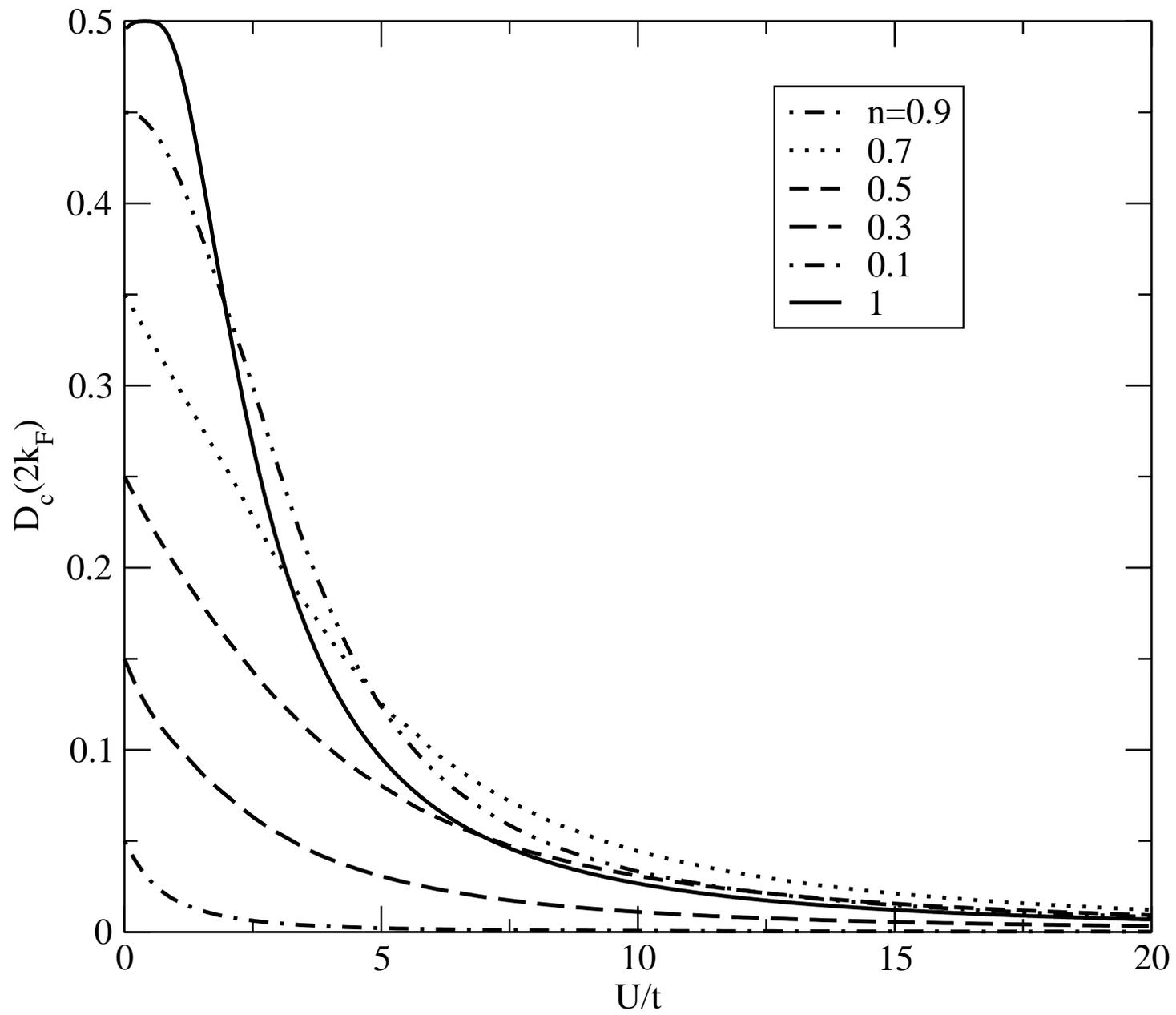
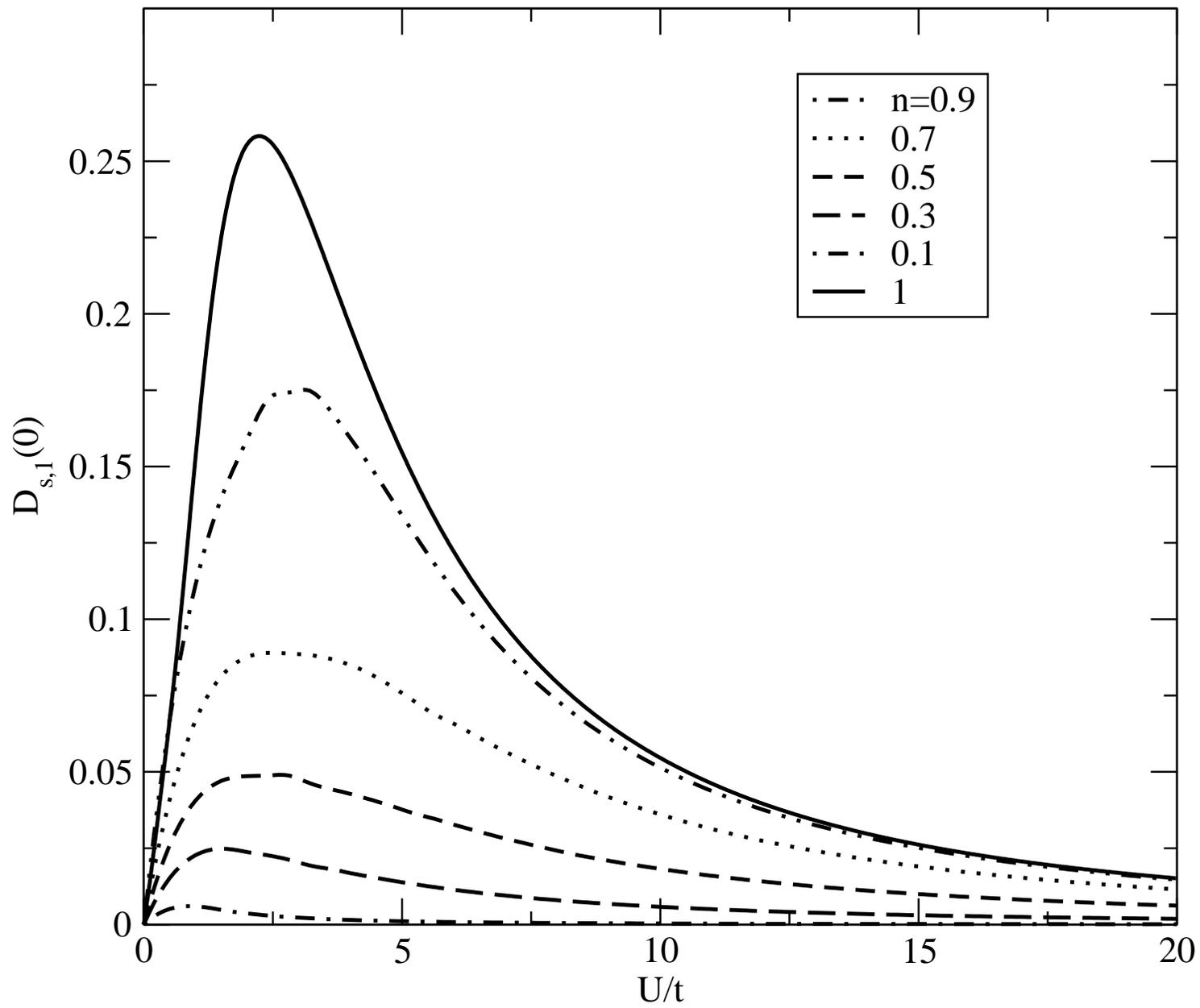
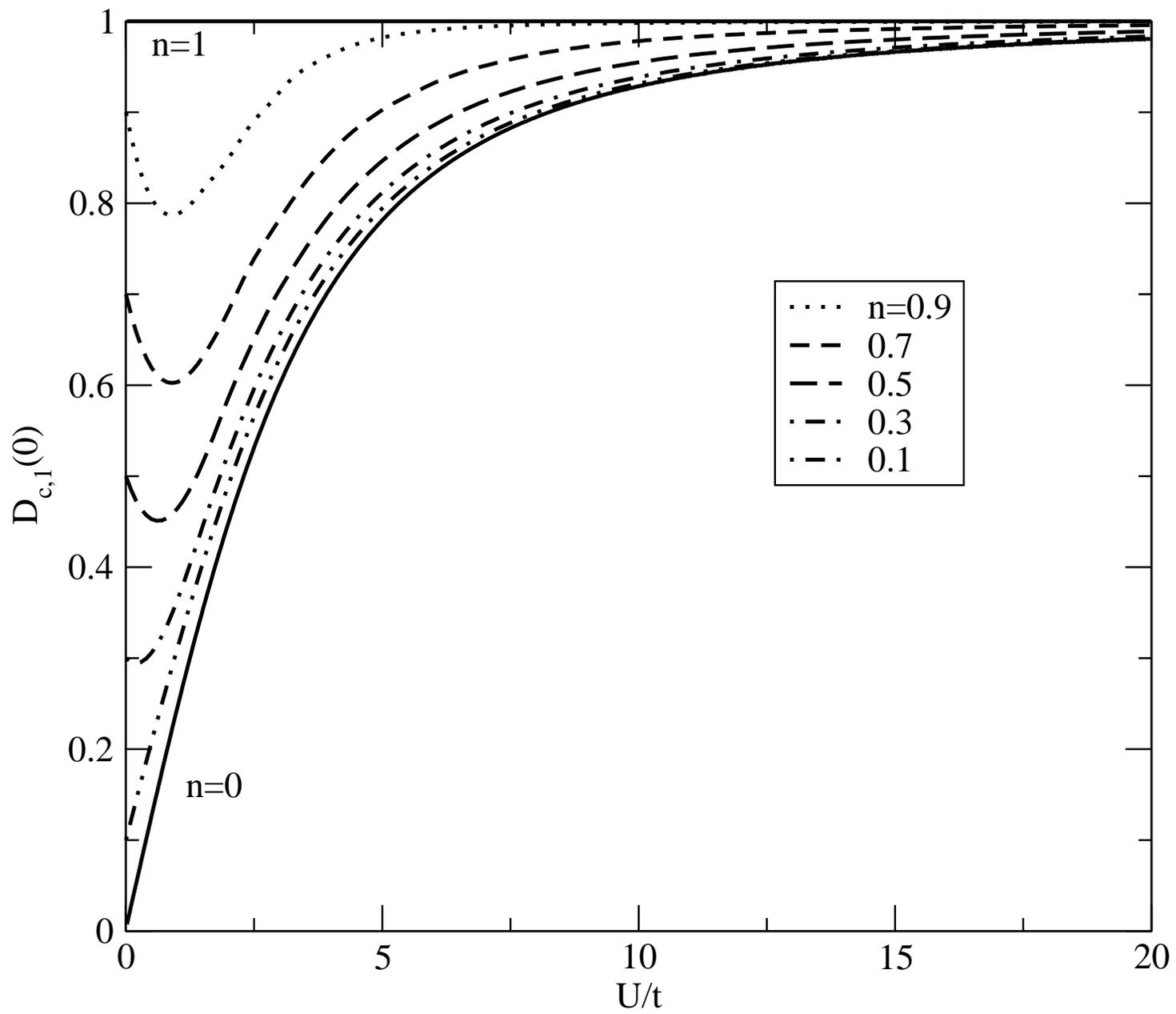
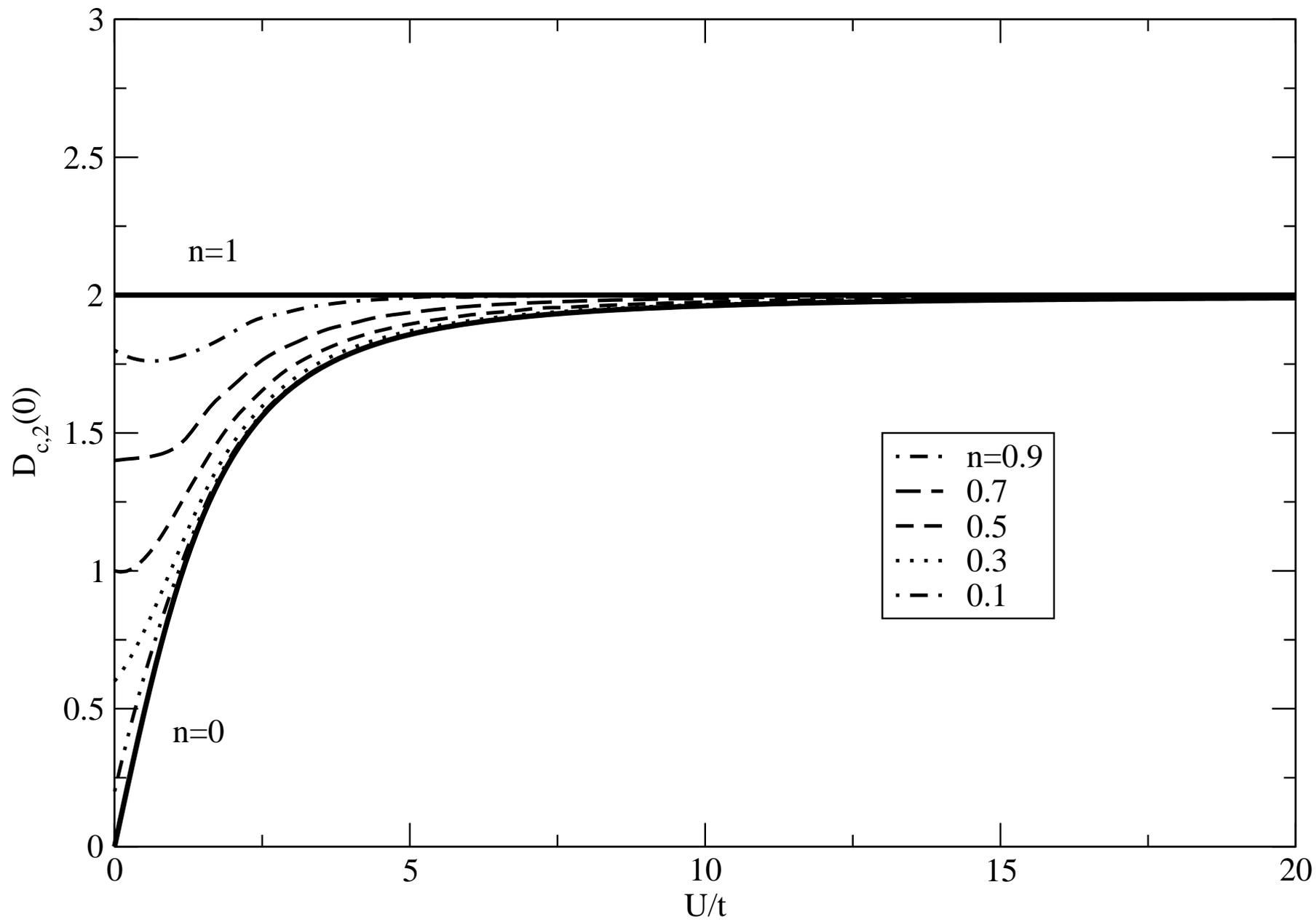
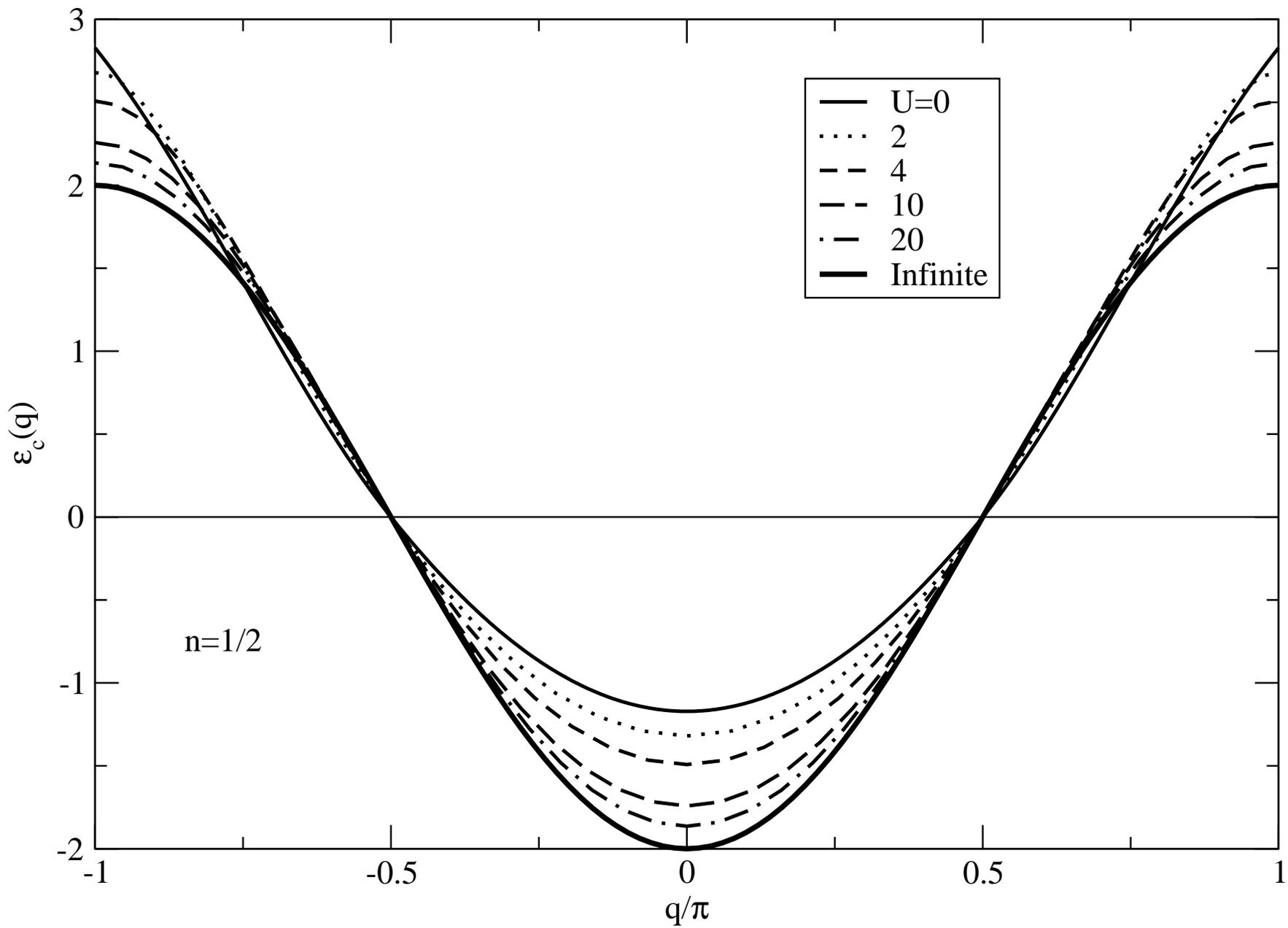
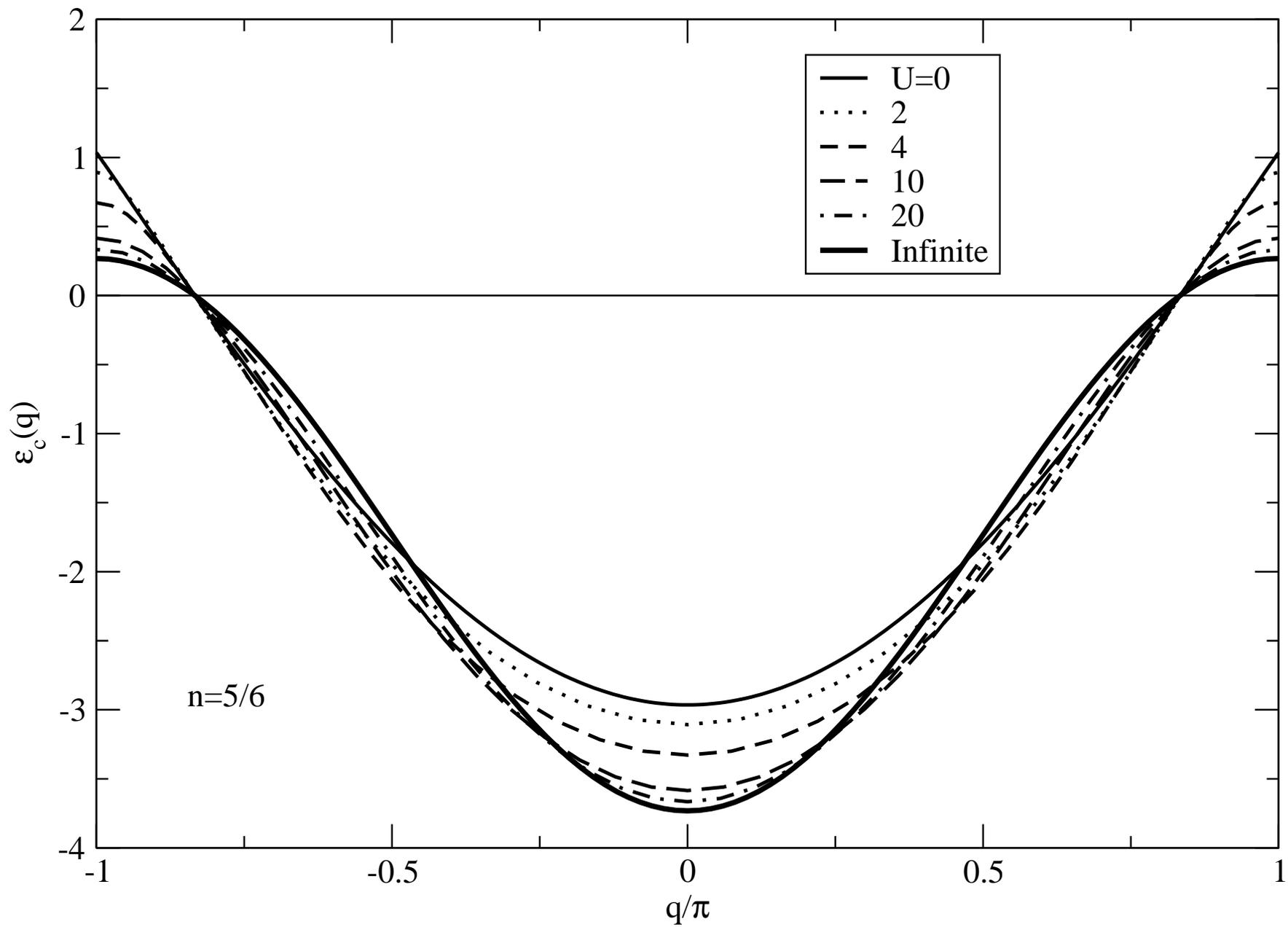
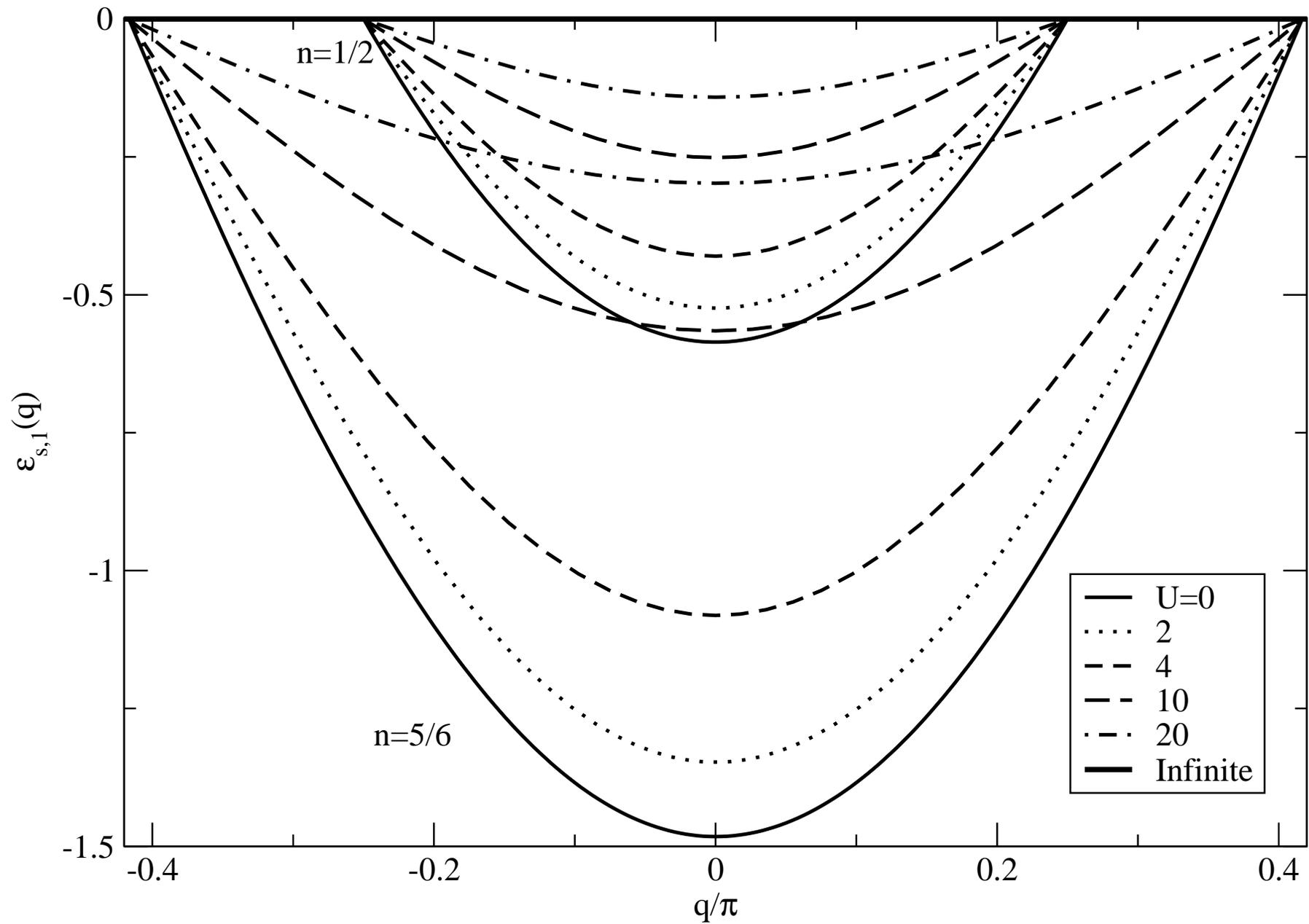
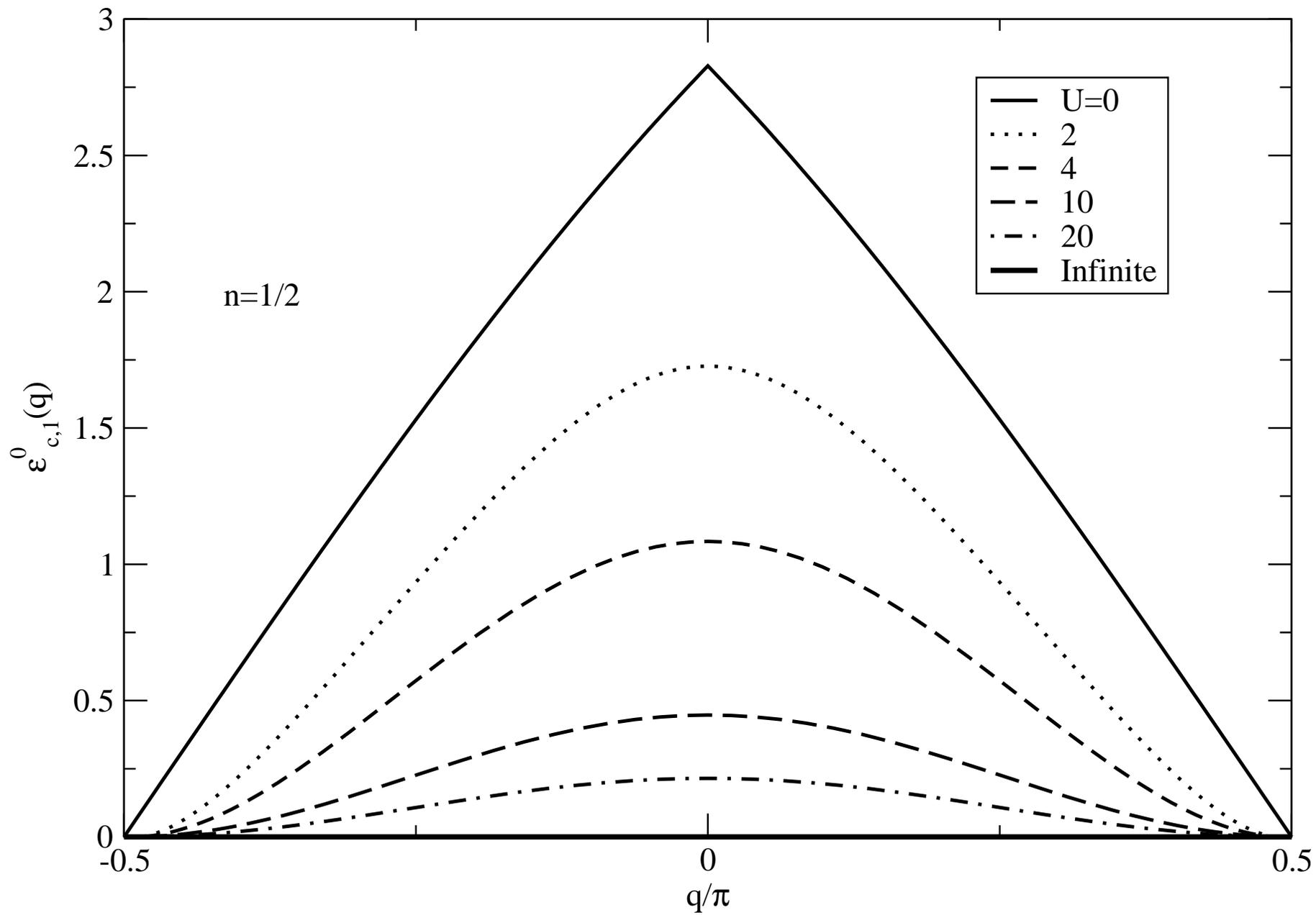
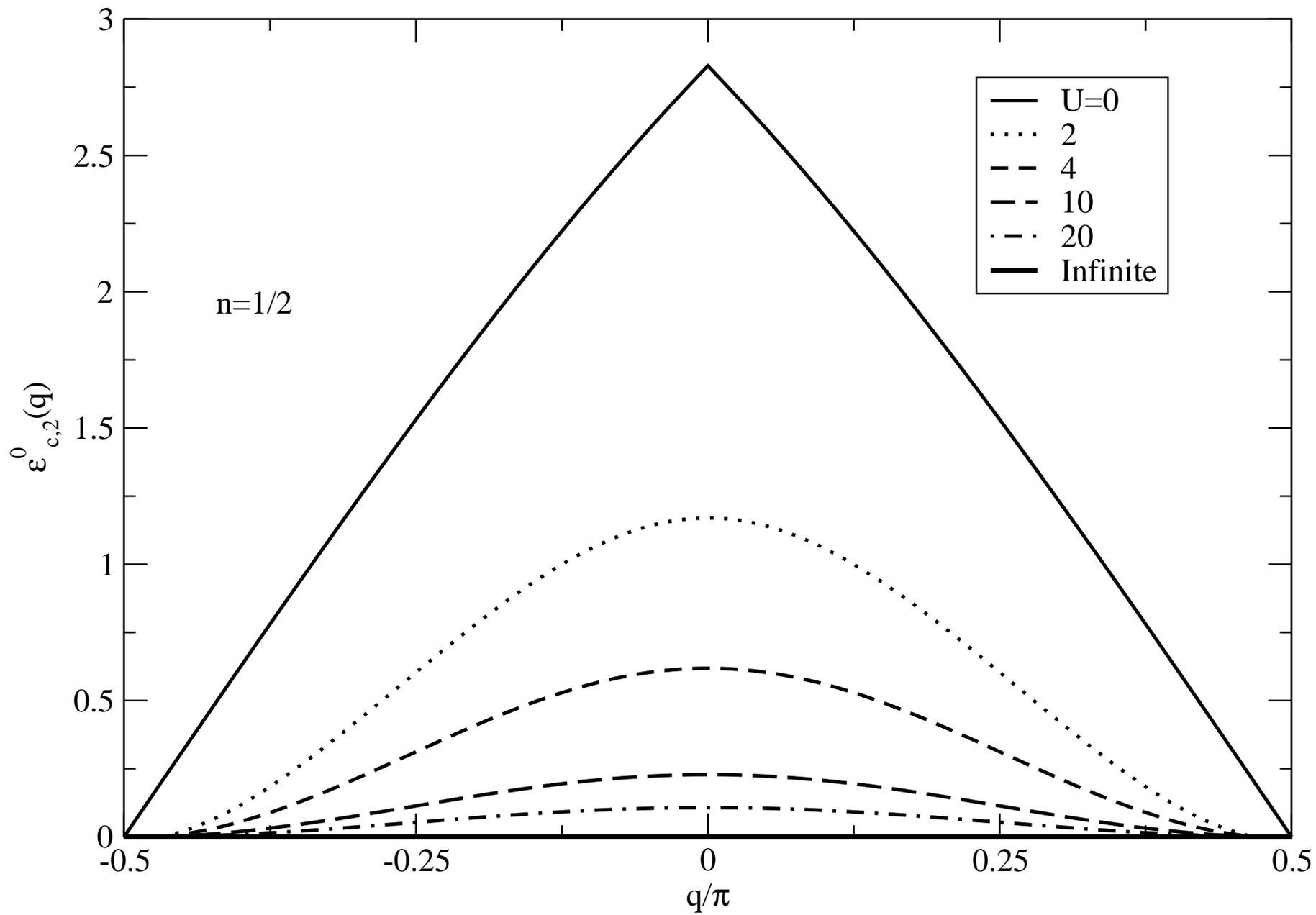
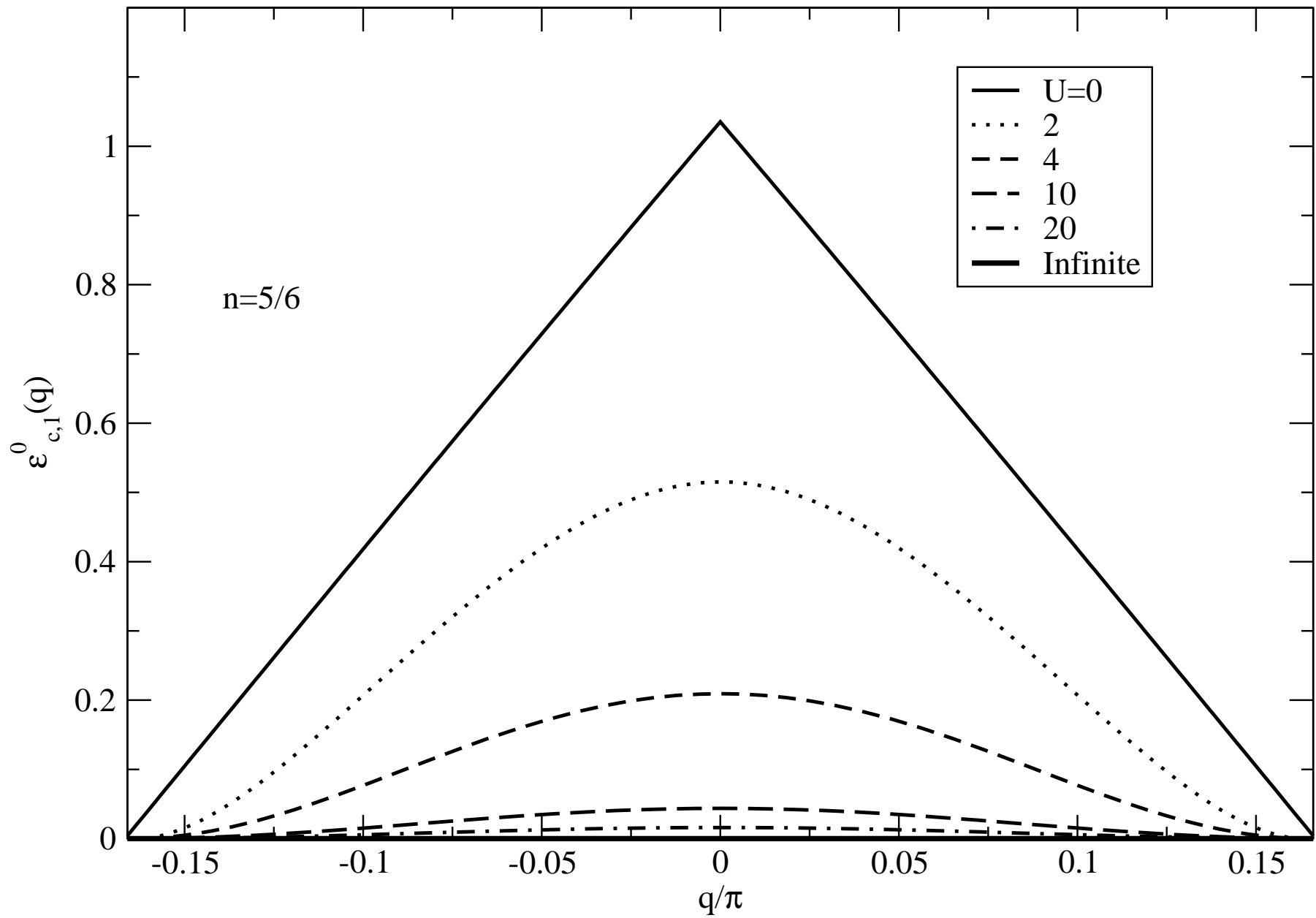
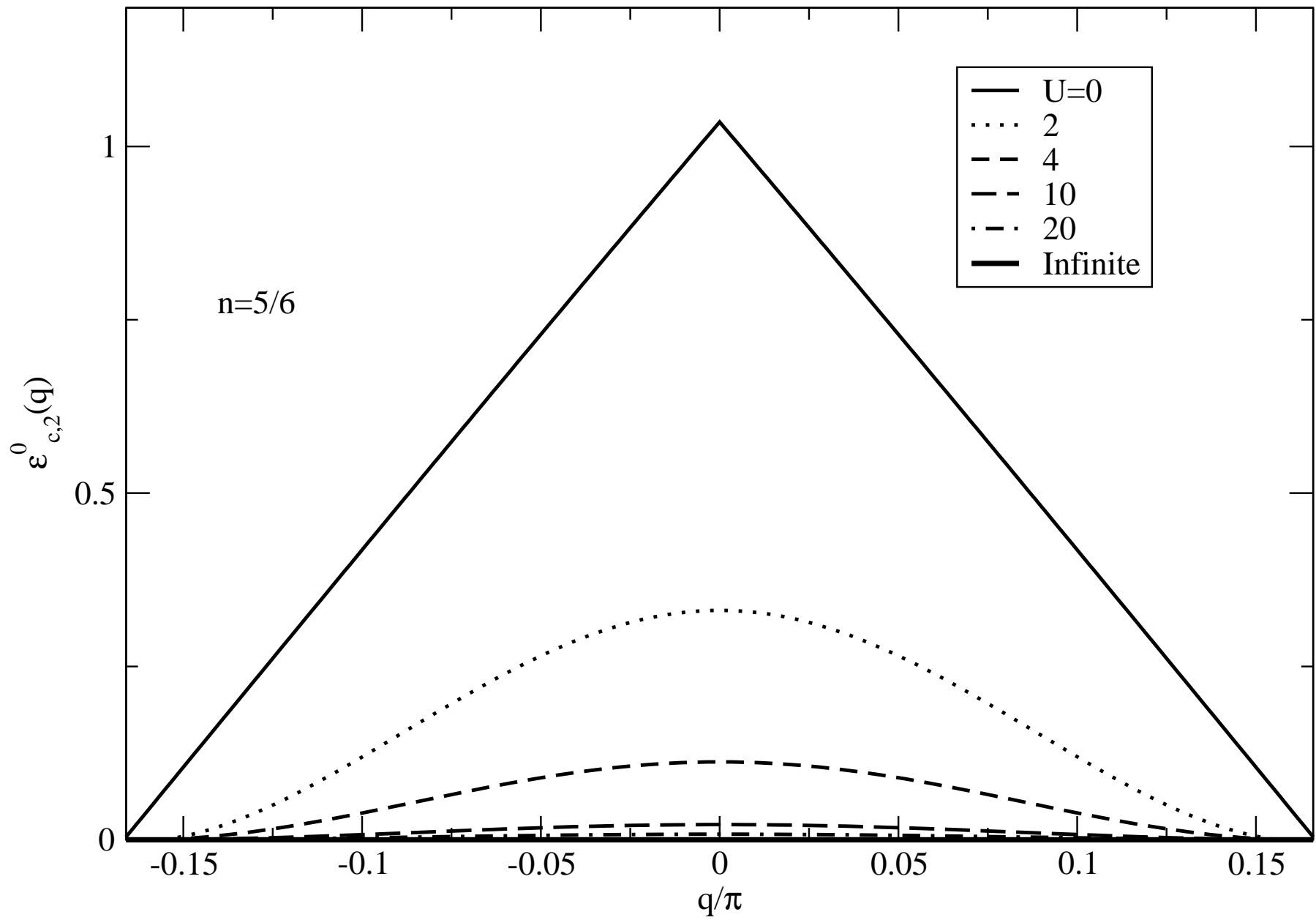
\end{document}